\pdfoutput=1
\documentclass{aa}  
\def\lofar{{\em LOFAR}}

\def\rosat{{\em ROSAT}}
\def\xmm{{\em XMM-Newton}}
\def\suzaku{{\em Suzaku}}
\def\asca{{\em ASCA}}

\usepackage{amssymb}
\usepackage{mathtools}
\usepackage{natbib}
\usepackage{graphicx,url,multicol,enumitem,sidecap}
\usepackage[margin=20pt]{subfig} 
\usepackage{multirow} 

\usepackage{graphicx}
\usepackage{txfonts}
\begin{document}

   \title{A low-frequency view of mixed-morphology supernova remnant VRO 42.05.01, and its neighbourhood}
   \author{M.~Arias \inst{1} \and
    J.~Vink \inst{1,2,3} \and
    M.~Iacobelli \inst{4} \and 
    V.~Dom{\v{c}}ek \inst{1,2} \and
    M.~Haverkorn \inst{7} \and 
    J.~B.~R.~Oonk \inst{4,5,8} \and \\
    I.~Polderman \inst{7} \and
    W.~Reich \inst{6} \and
    G.~J.~White \inst{9,10} \and
    P.~Zhou \inst{1} 
          }

   \institute{
   Anton Pannekoek Institute for Astronomy, University of Amsterdam, Science Park 904, 1098 XH Amsterdam, The Netherlands \and
   GRAPPA, University of Amsterdam, Science Park 904, 1098 XH Amsterdam, The Netherlands \and 
   SRON, Netherlands Institute for Space Research, Utrech, The Netherlands \and
   ASTRON, Netherlands Institute for Radio Astronomy, Postbus 2, 7990 AA, Dwingeloo, The Netherlands \and
   Leiden Observatory, Leiden University, P.O. Box 9513, NL-2300 RA Leiden, The Netherlands \and
   Max Planck Institute for Radio Astronomy, Auf dem H\"ugel 69, 53121 Bonn, Germany \and
   Department of Astrophysics / IMAPP, Radboud University Nijmegen, PO Box 9010, 6500 GL Nijmegen, The Netherlands \and
   SURFsara, P.O. Box 94613, 1090 GP Amsterdam, The Netherlands \and
   Department of Physics and Astronomy, The Open University, Walton Hall, Milton Keynes, MK7 6AA, UK \and
   RAL Space, STFC Rutherford Appleton Laboratory, Chilton, Didcot, Oxfordshire, OX11 0QX, UK
             }

   \date{Received XX; accepted XX}

 
  \abstract
   {Mixed-morphology supernova remnants (MM SNRs) are a mysterious class of objects that display thermal X-ray
   emission within their radio shell. They are an older class of SNRs, and as such are profoundly affected by the environment
   into which they evolve. VRO 42.05.01 is a MM SNR of puzzling morphology in the direction of the Galactic anticentre.}
   {Low-frequency radio observations of supernova remnants are sensitive to synchrotron electrons accelerated in the shock front. 
   We aim to compare the low-frequency emission 
   to higher frequency observations to understand the environmental and shock acceleration conditions that have given rise to the observed properties of this source.}
   {We present a LOFAR High Band Antenna map centred at 143 MHz of the region of the Galactic plane centred at $l = 166\degr, \,b = 3.5\degr$ at 143 MHz, with a resolution of 
   148\arcsec and an rms noise of 4.4 mJy\,bm$^{-1}$ . Our map is sensitive to scales as large as 6\degr. We compared the LOw Frequency ARay (LOFAR) observations 
   to archival higher frequency radio, infrared, and
   optical data to study the emission properties of the source in different spectral regimes. We did this both for the SNR and for OA 184, an H II region 
   within our field of view.}
   {We find that the radio spectral index of VRO 42.05.01 increases at low radio frequencies; i.e. the LOFAR flux is higher than expected from 
   the measured spectral index value at higher radio frequencies. This observed curvature in the low-frequency end of the radio spectrum occurs
   primarily in the brightest regions of the source, while the fainter regions present a roughly constant power-law behaviour between 143 MHz and 2695 MHz.
   We favour an explanation for this steepening whereby radiative shocks have high compression ratios 
   and electrons of different energies probe different length scales across the shocks, therefore sampling regions of different compression ratios.}
   {}


   \maketitle
%

\section{Introduction}

The evolution of supernova remnants (SNRs) is greatly affected by their various environments, which become the dominant factor in the later stages of the life of a remnant.
Galactic SNRs are often radio-bright and have large angular scales, allowing us to conduct detailed morphological studies of the sources and their neighbourhood.
In particular, the line of sight of the Galactic anticentre towards the Perseus arm
is less absorbed and less crowded than a typical Galactic field, while still rich in extended spiral-arm features.
Supernova remnants in this direction  
are excellent sites to study the interaction of a SNR with its surrounding
 circumstellar and interstellar media. More specifically, the synchrotron emission
from their shock-accelerated, relativistic electrons
directly probes the acceleration processes that are taking place.


   \begin{figure}
   \centering
   \includegraphics[width=\hsize]{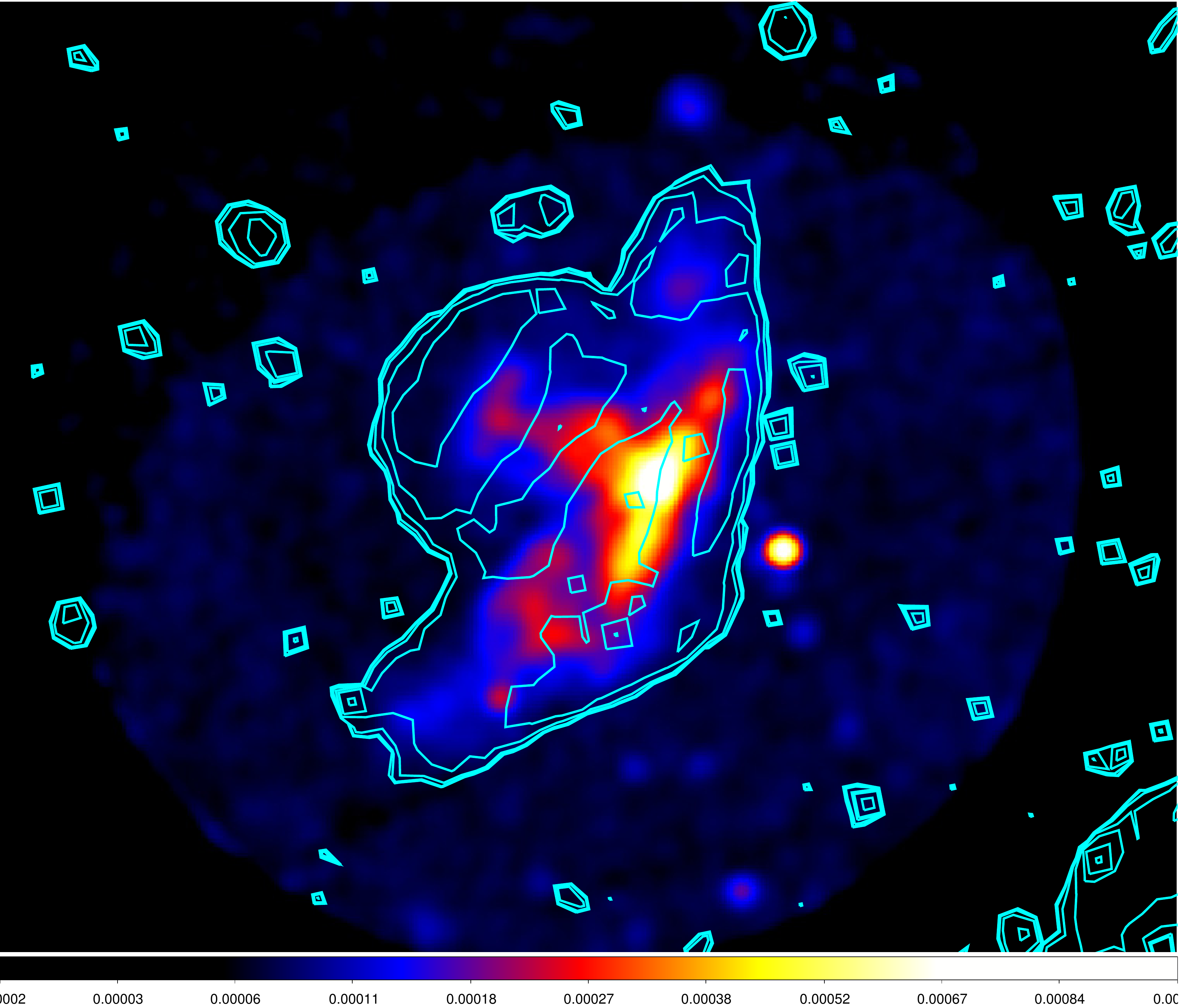}
      \caption{Exposure-corrected and smoothed X-ray mosaic of VRO 42.05.01 in the 0.1-2.4 keV energy band. The mosaic was constructed from archival ROSAT PSPC observations (RP500001N00 and RP500120N00). The cyan contours correspond to CGPS radio observations at 1420 MHz. The X-ray resolution is 1.25\arcmin and the radio resolution is 1.7\arcmin.
              }
         \label{x_rays}
   \end{figure}

\vspace{0.5cm}
\noindent
Mixed-morphology supernova remnants \citep[MM SNRs;][]{rho98} are a class of SNRs that display a shell-like morphology in the radio 
but are centrally dominated in the X-rays. Their X-ray emission is characterised by a thermal spectrum and is not due
to the presence of a pulsar wind nebula.
These objects are in a late stage of SNR evolution in which radiative energy losses become dynamically important. 
This happens when the post-shock temperature falls below $\sim 5 \times 10^5$ K, which 
corresponds to shock velocities $V_{s} \sim 200$ km s$^{-1}$, resulting in optical line emission that further 
cools the remnant \cite[e.g.][]{blondin98}. 
Mixed-morphology SNRs are typically older than 10,000 years, although there are a few young MM SNRs:
3C391 is $3,700-4,400$ years old \citep{chen01}, and 
W49B has been calculated to be, in one study, $1,000-4,000$ years old \citep{hwang00} and, in another study, $5,000-6,000$ years old \citep{zhou17}.
Moreover, the ages of some of some other MM SNRs have not been clearly determined: IC443 has
been proposed to be from $3,000-4,000$ years old to $20,000-30,000$ years
old \cite[see e.g.][]{troja08, gaensler06}, for example. 
Another characteristic of MM SNRs is that they tend to have a spectral index flatter than the
$\alpha = 0.5$ predicted by standard shock acceleration theory \cite[for $S_\nu \propto \nu^{-\alpha}$; c.f. table 4 in ][]{vink12}.
Many of them have been identified to be the result of a core-collapse explosion \cite[]{vink12}, although
some  might have been thought to be core-collapse remnants because of their mixed-morphology status, and in a
few cases a thorough study of the X-ray abundances has indicated that the MM SNR 
was in fact the result of a type Ia explosion \cite[][for remnants G344.7$-$0.1, G272.2$-$3.2, and W49B, respectively]{yamaguchi12, sezer12, zhou17}.

Remnants of the mixed-morphology class tend to be associated with the denser parts of the interstellar medium (ISM)
and often show signs of molecular interaction, as evidenced for instance
from OH maser emission \cite[e.g. IC443, W28, 3C391, etc.; see Table 4 in][]{vink12}. This is perhaps to
be expected if they are the remnants of the explosions of young massive stars in crowded environments, possibly
close to the molecular cloud(s) where the stars were born.
In fact MM SNRs are often bright GeV $\gamma$-ray sources and are the only SNRs for which proton cosmic rays 
have been unambiguously detected \cite[][for the SNRs W44 and IC 443]{ackermann13, giuliani11}.  
In X-rays the interiors of MM SNRs show evidence for enhanced abundances \cite[]{lazendic06b}, indicating the 
presence of supernova ejecta in addition to the shocked ISM material, 
and their X-ray emitting plasma tends to be overionised, indicating relatively high densities and rapid adiabatic expansion. 
These properties suggest that MM SNRs are important for understanding how shock interaction and cosmic-ray acceleration occur in dense environments.


   \begin{figure*}
   \centering
            {\includegraphics[width=0.99\textwidth]{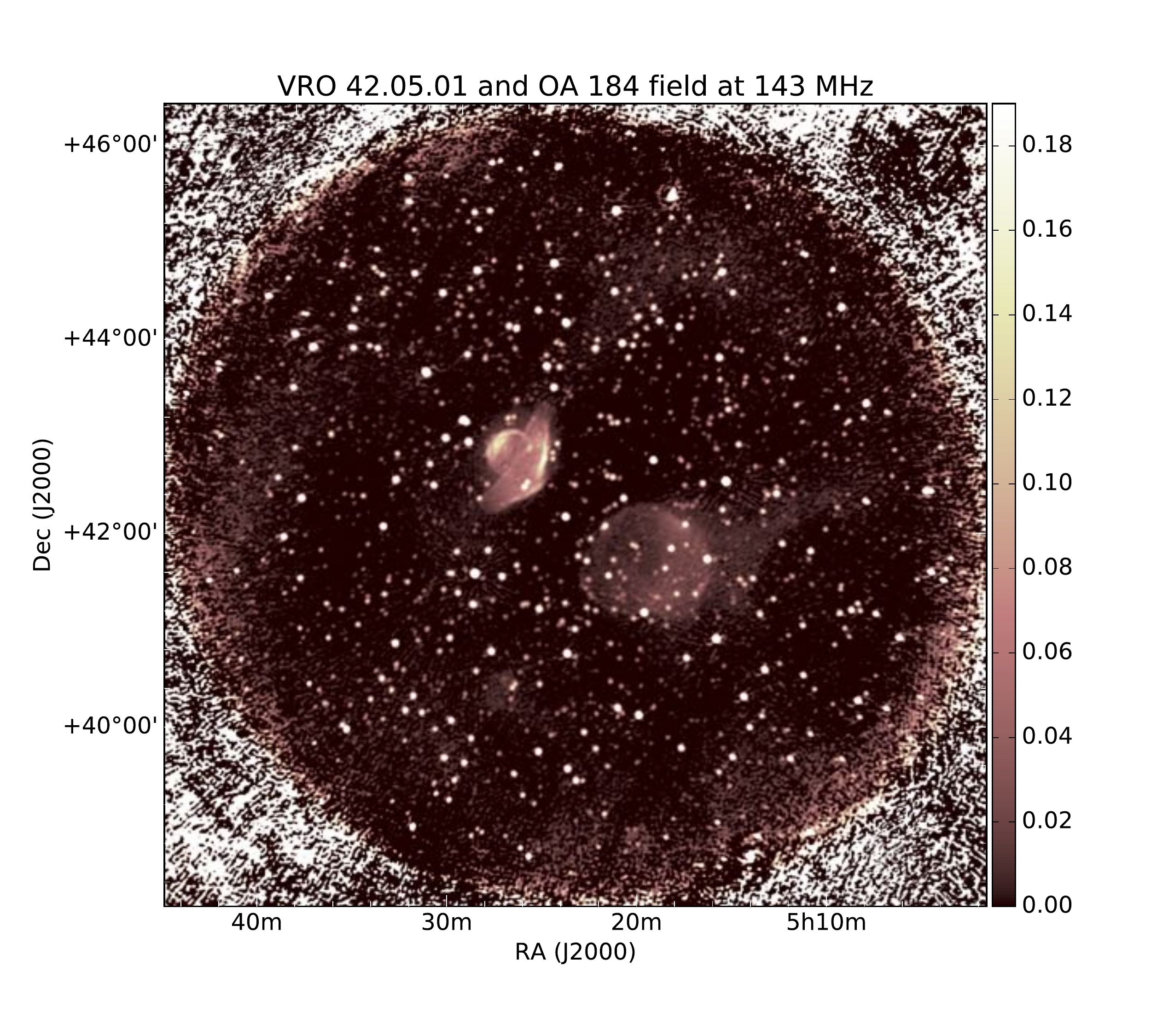}
      \caption{Full LOFAR field centred at VRO 42.05.01 and OA 184. 
      This image has a resolution of 148\arcsec\  and is made using only the core stations for best sensitivity to diffuse, extended emission. It has been primary-beam corrected. 
      The central frequency is 143 MHz and the bandwidth is 59 MHz. The noise at the centre of the image is 4.4 mJy~bm$^{-1}$; the colour scale to the right of the map is in Jy~bm$^{-1}$.
      VRO 42.05.01 is the two-shell-shaped extended structure on the left-hand side of the image; the smaller semicircle is the \lq shell' and the wider semicircle
      the \lq wing'.OA 184 is the nearly circular shell to the right.
      The trail of diffuse emission to the west of OA 184 is referred to in the text as the  synchrotron tongue.
              }
  \label{full_field}}    
   \end{figure*}

\vspace{0.5cm}
\noindent
VRO 42.05.01 (G\,166.0\,+4.3) is a MM SNR in the direction of the Galactic anticentre (RA=05:26:30, Dec=+42:56). 
It has a peculiar two-shell radio shape whose outline is filled in X-ray  
emission (see Figs. \ref{x_rays} and \ref{full_field}). This object displays filamentary optical line emission that is broadly coincident with the outline of the radio shell,
and line ratios that suggest a low-density environment
\cite[]{vandenbergh73, fesen85}. \cite{araya13} detected GeV $\gamma$-ray emission from the source, which, they argued, has a 
leptonic origin (unlike other MM SNRs). Two targeted pulsar searches were conducted in the region \cite[]{biggs96,lorimer98},
but no pulsar was found. 
Its distance has been estimated to be 4.5$\pm$1.5 kpc, the distance to the Perseus arm at the Galactic
latitude of $l=166$\degr \cite[]{landecker89}.

Our lack of understanding of the morphology of VRO 42.05.01 is twofold:
we lack a widely accepted model for the presence of dense X-ray emitting material in the centre of MM SNRs more generally; and 
it is not clear what conditions in the ejecta and surrounding medium could give rise to the peculiar shape of the source. 

The latter has been the subject of speculation for several decades. A series of papers
\cite[]{landecker82, pineault85, pineault87, landecker89} proposed that the supernova explosion had taken place in a dense medium,
and that a part of the expanding bubble eventually broke into a more tenuous medium, which allowed it to 
expand at a faster velocity. 
These papers refer to the smaller semicircle as \lq the shell', and the triangular larger component
as \lq the wing', a terminology that is maintained throughout the literature (these features are labelled
in Fig. \ref{regions}). 
Although abrupt density discontinuities in the ISM are common and this scenario is plausible, 
no unambiguous signs of interaction with the medium have been observed.
The relatively low surface density of VRO 42.05.01 (7 Jy at 1 GHz for an angular size of 45\arcmin$\times$80\arcmin) 
does not seem to agree with the idea that it is evolving in a 
high-density region.
Moreover, the CO survey of \cite{huang86} did not show any conspicuous molecular cloud in the face of the 
remnant, nor clear indications of sharply different densities. On the other hand, 
the remnant has a flat radio spectral index, $\alpha \approx 0.37$ \cite[]{leahy05}.
The subject of flat spectral indices in evolved remnants is discussed later in this paper in more detail, but
flat radio spectral indices might arise in high compression ratio shocks, suggesting a high post-shock density.

In the case of older, irregular, impossibly-shaped remnants the idea
that they might be two SNRs that occurred within a short time is often suggested.
The X-ray morphology of VRO 42.05.01, which is internal both to the shell and the wing,
 would require that both SNRs are MM even though only $\sim$10\% of the Galactic SNR population is MM \cite[]{vink12}.
The triangular shape of the larger shell, which has an abrupt ending in a flat line on the east, also disfavours 
this scenario. 
However, the possibility that VRO 42.05.01 might be two SNRs has not
been completely ruled out.

VRO 42.05.01 has been observed at X-ray wavelengths with \rosat\ \cite[]{burrows94}, \asca\ \cite[]{guo97},
\xmm\ \cite[]{bocchino09}, and \suzaku\ \cite[]{matsumura17}. Both \cite{burrows94} and \cite{guo97}
found a temperature of $\sim 8.3 \times 10^{6}$ K in the X-ray emitting material and remarked on the unusually low column density 
towards the remnant ($2.9 \times 10^{21}$ cm$^{-2}$). \cite{bocchino09} found that the abundances are well described by
a thermal equilibrium model, but the \suzaku\
observations in \cite{matsumura17} are consistent with the presence of a recombining plasma in the smaller shell. 
The authors proposed that this might be due to
heat being conducted from the SNR to a molecular cloud that allegedly sits at the north of the remnant, 
which lowers the temperature of the SNR. The plasma, they argued, is overionised, as it still has not reached the
lower ionisation states associated with cooler temperatures.


\section{Data reduction}

\subsection{Observations, calibration, and imaging}

 \begin{table}
\caption[]{Table of observations.}
\label{table_obs}     
\centering                          
\begin{tabular}{l c }   
\hline                
Number of stations & 22 \\ 
Central frequency & 143 MHz \\ 
Bandwidth & 59 MHz \\
Resolution & 148\arcsec \\
Min baseline & 10 m \\
Max baseline & 4 km \\
\hline    
\end{tabular}
\end{table}

We took eight hours of LOFAR High Band Antenna (HBA) data in January 2016 under project LC5\_012. 
Fifteen minutes at the beginning and the end of the observation were spent on calibrator 3C147,
and the remaining time was spent on the field of VRO 42.05.01. 
For these observations the HBA Dual station configuration was adopted, which is not optimal for calibration and imaging.
For this reason, only the visibilities from the core stations were selected and used in the processing.
The data were flagged for radio frequency interference, and the data from stations 1 and 13 were removed for the calibrator
and the target because they were very noisy.

The 5 Jy source 3C134 was in the edge of our field of view, approximately 5 degrees from the phase centre. 
A bright source like this one causes strong artefacts in a large portion of the image, and so we peeled the source
\cite[i.e. removed its contribution from the visibilities;][]{noordam04} with the \texttt{SAGECAL} algorithm \cite[]{yatawatta08}. 

The calibrator was calibrated in a direction-independent manner using the Pre-Facet Calibration Pipeline \cite[]{vanweeren16}. The pipeline obtains diagonal solutions towards the calibrator, 3C147, and then
performs clock-TEC separation, which distinguishes between clock offsets and drifts, and signal delays due to the electron column density in the ionosphere. Finally, this pipeline transfers the 
clock corrections and calibrator gain amplitudes to the target data set.

The visibilities were imaged using the \texttt{wsclean} algorithm \cite[]{offringa14}. The full bandwidth of the LOFAR HBA was used for
making Fig.~\ref{full_field}, which was imaged in the multi-scale setting and with a Briggs weighting robust parameter of $-2$ \cite[]{briggs95}.
The uniform weighting scheme is possible in the context of these observations because for the LOFAR core stations
the baseline distribution is smooth.
We imaged using only the LOFAR core stations, resulting
in a beam size of 148\arcsec. This allowed us to recover the degree-scaled extended emission present in the field.
As can be seen in the figure, some bright regions in the extended sources have negative bowls around them owing to the lack of zero spacings.


We pointed the telescope at  RA=05:23:15, Dec=+42.25.45,
centred between the two large extended sources in the field: the double-shell-shaped VRO 42.05.01 and the
H II region OA 184, which is a large ellipse of semimajor axis 80\arcmin\ and semiminor axis 75\arcmin. In addition, in the map we can see some large-scale synchrotron emission parallel to the Galactic plane
(see Fig. \ref{vel_plots} to see the orientation of the sources in Galactic coordinates).
The tail of emission at the west of OA 184, which extends 80\arcmin\ similar to a comet-tail,
was discussed in \cite{leahy05} and referred to as a \lq synchrotron tongue' from observations at 408 and
1420 MHz. There are two similar tails to either side of the wings of VRO 42.05.01, which are in fact also visible at those higher frequencies \cite[see e.g. the full resolution
1420 MHz map in Fig. 1 of][]{leahy05}.

\subsection{Total flux density calibration}

   \begin{figure}
   \centering
   \includegraphics[width=\hsize]{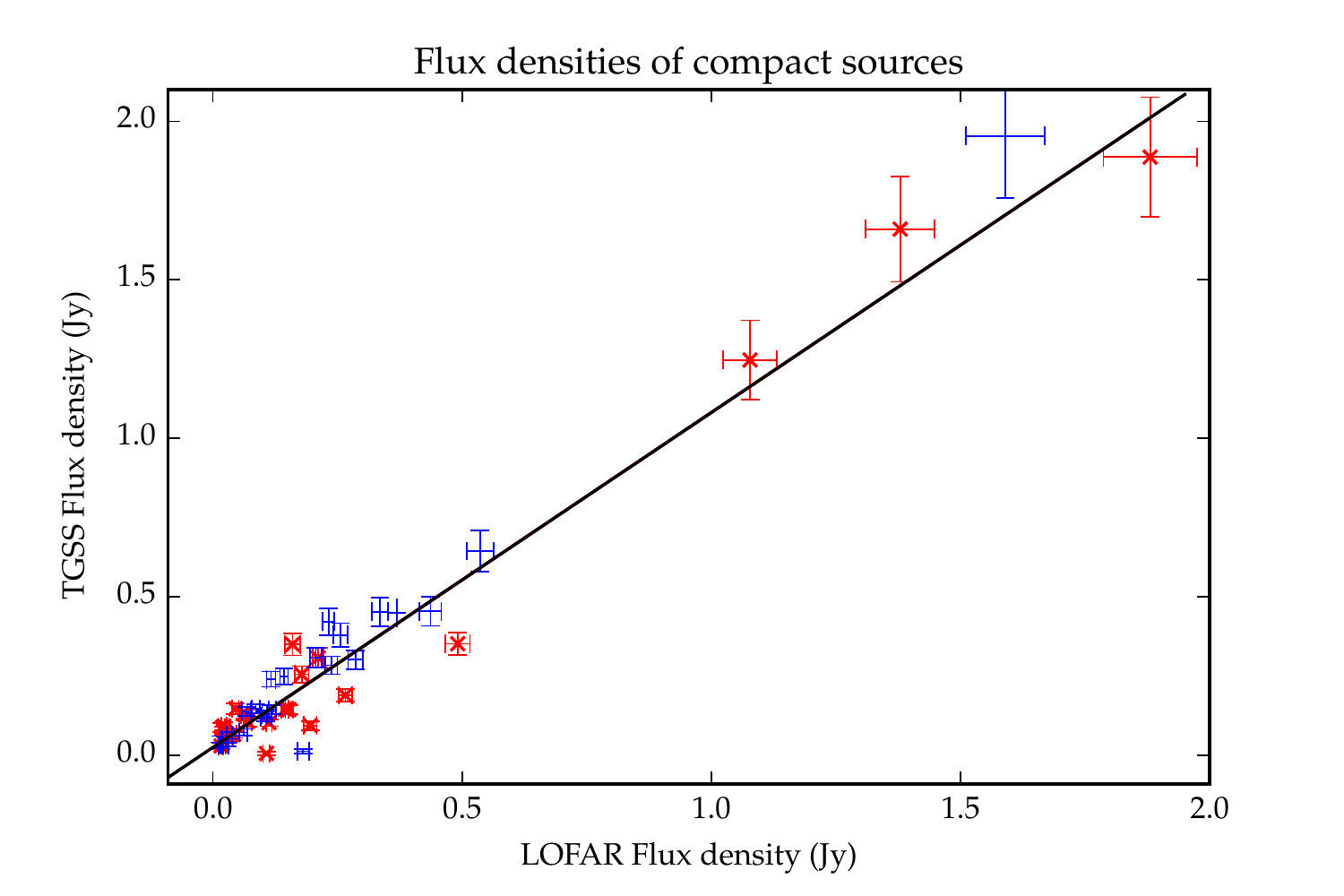}
      \caption{Comparison of flux density values from the LOFAR HBA images at 143 MHz and the TGSS image at 147 MHz of unresolved sources nearby VRO 42.05.01 (red) and
      OA 184 (blue). The best-fit line in black has slope $1.05 \pm 0.08$ and intercept $0.025 \pm 0.007$.
              }
         \label{flux_dens_comp}
   \end{figure}

As can be seen from Fig. \ref{full_field}, the noise of the image is non-uniform, in part because of the LOFAR beam and in part because of the unmodelled large-scale emission from the Galactic plane. 
For this reason, we picked $\sim50$ unresolved sources in the vicinity of VRO 42.05.01 and OA 184 and compared their flux densities in the full bandwidth map at 143 MHz
to their flux density in the TIFR GMRT Sky Survey catalogue at 147 MHz \cite[TGSS;][]{intema17}. 
The flux densities as obtained from the HBA and TGSS images are plotted in Fig. \ref{flux_dens_comp}. 

As one can see from this plot, LOFAR measures flux density values for these compact sources that are systematically lower than those of TGSS by a factor of 0.95. 
Differences in flux densities measured by LOFAR and the GMRT have been noted in \cite{vanweeren16} and \cite{shimwell17}. 
Our small difference of a factor of 0.95 is well within the uncertainty in the calibration of the low-frequency flux scale
with the GMRT \cite[estimated to be around 10\%,][]{intema17}. 
Moreover, the systematic calibration errors in the LOFAR flux scale are of the order of 10\%,
which can explain the difference.
We note that we are comparing two nearby but not identical
frequencies (143 MHz for LOFAR and 147 MHz for the GMRT). Rescaling the LOFAR flux densities to the TGSS
frequency with a spectral index of 0.8, typical for extragalactic sources, makes the LOFAR flux densities systematically lower than the TGSS flux densities by a factor of 0.93 instead of 0.95,
which is still within the uncertainty in the low-frequency flux scale.

 \begin{table}
\caption[]{Flux densities, LOFAR in-band spectral index, and LOFAR-CGPS spectral index, with and without compact source (CS) subtraction.
For comparison we also include the spectral indices from \cite{leahy05}.}
\label{fluxes_table}     
\centering                          
\begin{tabular}{l c c}        
\hline                 
Source & VRO 42.05.01 & OA 184  \\    
\hline                        
$S_{143 \mathrm{MHz}}$ with CS & $16.4\pm0.1$\,Jy &  $18.3\pm0.1$\,Jy \\      
$S_{143 \mathrm{MHz}}$ CS subtracted& $15.8\pm0.1$\,Jy & $11.2\pm0.2$\,Jy  \\
\hline
In-band $\alpha$ with CS &    $0.58\pm0.04$ & $0.48\pm0.02$  \\
In-band $\alpha$ CS subtracted& $0.57\pm0.04$ &  $0.37\pm0.02$  \\
\hline                                   
$\alpha$ LOFAR-CGPS with CS & $0.49\pm0.02$ &  $0.32\pm0.01$  \\
$\alpha$ CS subtracted& $0.49\pm0.02$  &  $0.16\pm0.01$  \\
\hline    
$\alpha$ L\&T05 with CS & $0.38 \pm 0.03$ & $0.28 \pm 0.10$ \\
$\alpha$ L\&T05 CS subtracted & $0.36 \pm 0.03$  & $0.21 \pm 0.12$ \\
\hline
\end{tabular}
\end{table}

In Table \ref{fluxes_table} we present the flux density values for VRO 42.05.01 and OA 184
as measured with LOFAR, as well as the in-band spectral index for each source (see the next section).
Both of the extended sources are covered in point sources. We present the flux density measurements for each source
both subtracting the point sources that are visible in the face of the extended source and without this subtraction, and we do the same thing for the spectral index. 
In the case of VRO 42.05.01 
we subtract the two bright sources evident in the wing. The third remaining point source over the remnant discussed in \cite{leahy05} 
is confused at LOFAR frequencies within the  extended emission of the remnant. In the case of OA184 we subtract the 26 sources above 14 mJy\,bm$^{-1}$ ($\sim3\sigma$ of the image noise)
that are visible in the face of the source.

\subsection{Narrow-band images and in-band spectral index}
\label{sec_spx}

\begin{figure*}
\centering
\includegraphics[width=\textwidth]{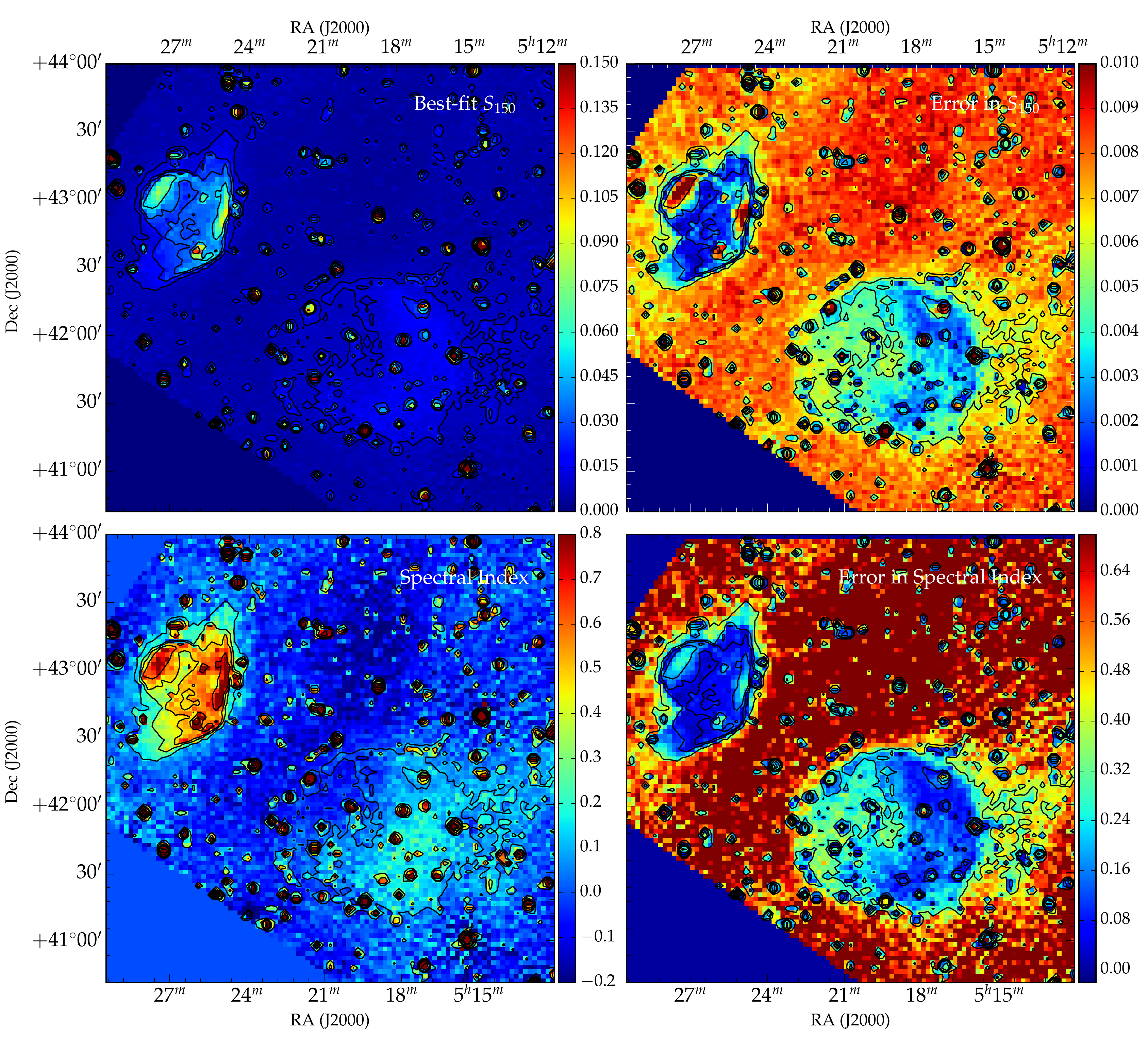}
\caption{Results of fitting LOFAR, CGPS, and Effelsberg data to 
Eq. \ref{eqn_pl}. Top left is the best-fit amplitude $S_{150\mathrm{MHz}}$ and top right is
the error in best-fit amplitude $\Delta S_{150\mathrm{MHz}}$. Bottom left is the best-fit spectral index
value $\alpha$ and right is the error in spectral index $\Delta\alpha$. 
The contours are those of the best-fit amplitude $S_{150\mathrm{MHz}}$.
}
\label{fit_results_all}
\end{figure*}
  
Since our HBA observations covered a large bandwidth (from 115 MHz to 174 MHz), we can obtain spectral information from our pointing. 
Shimwell et al. 2018 (submitted) have discussed that it is possible to
divide the LOFAR HBA bandwidth into three parts, make a three channel image, and accurately derive 
(with 10\% uncertainty or less) in-band spectra for sources with integrated flux densities $\geq5$ mJy 
(for maps with a typical rms of 0.1 mJy\,bm$^{-1}$).

For that reason, we divided the bandwidth of our data to make three  20 MHz wide images centred at 125, 144, and 164 MHz (see Fig. \ref{three_freqs}). 
These were made with a common uv-range of 10$\lambda$ to 1400$\lambda$. 

We report the value of the integrated spectral indices of the extended sources, both with and without compact source subtraction, in Table \ref{fluxes_table}.
To calculate these we fit a power law to the flux density values at the three frequencies.
We note that the in-band spectral index is calculated from three frequencies at 125, 144, and 164 MHz, and so it does not have much of a leverage arm in frequency 
and is severely subject to LOFAR systematics. 
We therefore also calculated the spectral index
including the flux densities from the Canadian Galactic Plane Survey \cite[CGPS; ][]{taylor03} images as reported in \cite{leahy05}. In this case we fitted the flux density values in Table \ref{fluxes_table}
and the 408 MHz and 1420 MHz values in Table 3 of \cite{leahy05} to a power law.

The spectral index values that we find are different to those in \cite{leahy05}, which were calculated between 408 MHz and 1420 MHz. The only value that agrees within 
the error bars is 
the spectral index of OA 184 without compact source subtraction. This is in fact surprising, as extragalactic compact sources typically
have $\alpha \sim0.7 - 0.8$ \cite[]{mahony16, williams16}, and so one would expect their contribution to steepen the overall spectrum at LOFAR frequencies. If we subtract the
compact sources, our value of $\alpha = 0.16\pm0.01$ is in fact smaller than that found in \cite{leahy05}.
For an optically thin HII region, $\alpha = 0.16$ is a reasonable spectral index value and part of the
discrepancy with \cite{leahy05} might be due to our different threshold values for removing the compact sources:
they remove 17 sources with flux densities as low as 20 mJy at 408 MHz, whereas we remove 26 sources above 14 mJy
at 143 MHz.

On the other hand, we find VRO 42.05.01 to have a significantly steeper index than previously reported. The LOFAR in-band spectral 
index ($\alpha$=0.58 and 0.57 with and without compact source subtraction, respectively) 
is steeper than the spectral index if we include the higher frequency flux densities ($\alpha$=0.49 and 0.49), 
and both are much steeper
than the spectral index found only from the CGPS data (0.37 and 0.36). 
We explore several scenarios that could account for an intrinsic steepening of VRO 42.05.01 
at LOFAR frequencies in section \ref{sec_interpretation}.

We made an in-band spectral index map from these three narrow-band images. It is presented in Fig. \ref{all_spxs}, upper left.
We note that the LOFAR in-band spectral index map is not trustworthy since the side lobes are not calibrated
as a function of elevation. The primary beam can also introduce spurious effects that affect the in-band spectral 
index radially. In addition to these instrumental effects, the large fluctuations in the in-band spectral index are 
also affected by artefacts in the narrow-band images (such as bowls around the bright sources).
The oscillations are of course also larger in the areas of poorer signal to noise. 
For these reasons, the backbone of our analysis is comparing the LOFAR map to higher frequency observations.

\subsection{Other data and higher frequency spectral index map}

In this paper we compare our LOFAR observations to  higher frequency archival data. 
At radio frequencies, 
we use the 408 MHz and 1420 MHz continuum maps from the CGPS \cite[][]{taylor03} 
as well as the 2695 MHz maps from the Effelsberg 11 cm Survey in the direction of the Galactic anticentre \cite[]{furst90}.
 The Effelsberg 11 cm data has a resolution of 4.3\arcmin. 
The CGPS samples baselines from 12.9 to 604.3 m, which correspond to scales of 1.7\arcmin$-56$\arcmin at 1420 MHz, and 
4.1\arcmin$-195$\arcmin at 408 MHz. This survey also includes single-dish data to account for zero spacings. 
For comparison, our LOFAR images are sensitive to scales of $\sim3$\arcmin$-360$\arcmin. VRO 42.05.01 and OA184 present substantial amount of
 diffuse emission on degree scales, but within the scales probed by LOFAR. 
 Therefore, the LOFAR, CGPS, and Effelsberg data can be safely compared.

We smoothed the 143 MHz, 408 MHz, and 1420 MHz to the 4.3\arcmin resolution of the 2695 MHz Effelsberg map and
made a spectral index map by  fitting each pixel in each of the images
to a power law of the following form:
\begin{equation}
S_\nu = S_{150\mathrm{MHz}} \left(\frac{\nu}{150\mathrm{MHz}}\right)^{-\alpha},
\label{eqn_pl}
\end{equation}
where our fit parameters were the flux density at 150 MHz $S_{150\mathrm{MHz}}$ and the spectral index $\alpha$. 
The best-fit spectral index, flux densities, and errors in each for VRO 42.05.01 and OA184 are plotted in Fig. \ref{fit_results_all}.


In addition to the continuum data products, we use 21 cm data cubes from the CGPS to study the neutral hydrogen environment of these 
sources. 
At infrared wavelengths we use data from the Wide-Field Infrared Survey Explorer all-sky survey \cite[WISE; ][]{wright10}.
Finally, we took the optical H$\alpha$ map from the Middlebury Emission-Line Atlas of Galactic SNRs, found 
at \protect\url{http://sites.middlebury.edu/snratlas/g166-04-3-vro-42-05-01/} (Winkler et al., in preparation). 

   \begin{figure}
   \centering
   \includegraphics[width=\hsize]{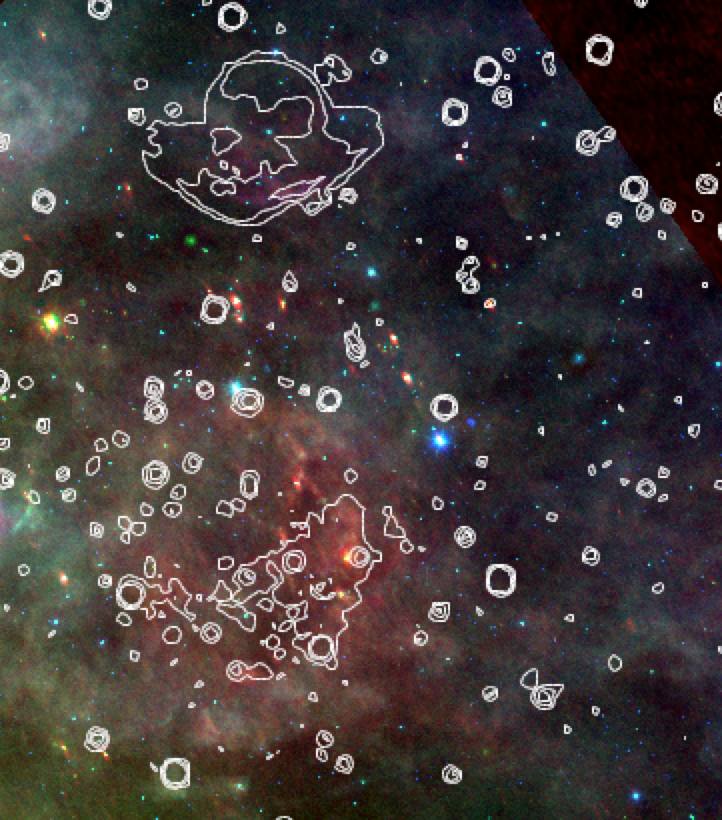}
      \caption{Three colour image made from IRAS $60 \, \mu$m image (red), Wise $22 \, \mu$m (green) and Wise $12 \, \mu$m (blue). 
      The contours are from the LOFAR full bandwidth image. 
              }
         \label{wise_rgb}
   \end{figure}

\subsection{Curvature in the radio spectrum of VRO 42.05.01}

 \begin{figure*}
\centering
\includegraphics[width=0.9\textwidth]{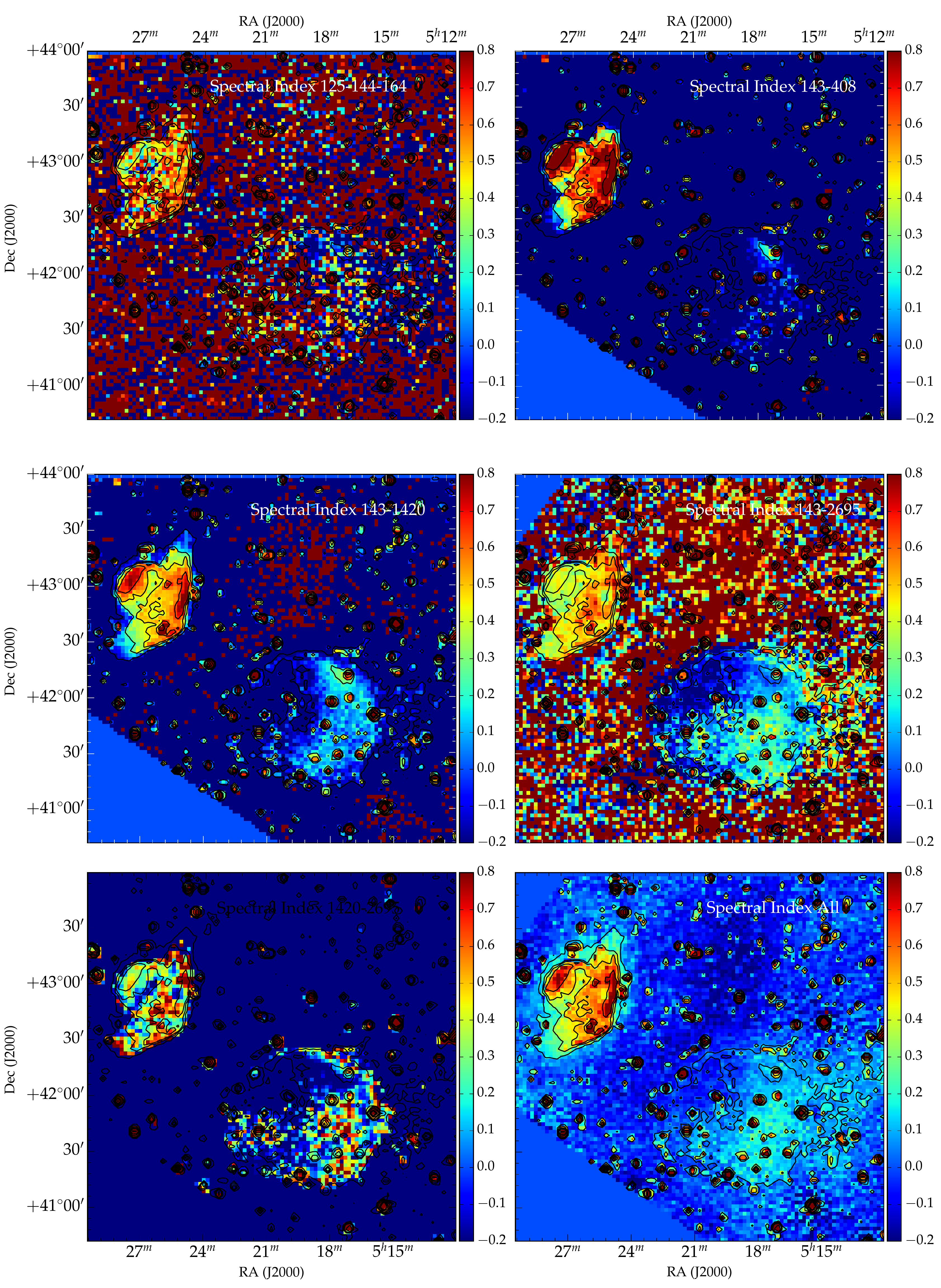}
\caption{Spectral index maps made from the LOFAR, CGPS, and Effelsberg data. \textit{Top left:} LOFAR in-band spectral
index made from the three narrow-band images described in section \ref{sec_spx} and presented in the appendix. 
\textit{Top right:} 143 MHz to 408 MHz (LOFAR to CGPS) spectral index. \textit{Middle left:} 143 MHz to 1420 MHz (LOFAR to CGPS) spectral index.
\textit{Middle right:} 143 MHz to 2695 MHz (LOFAR to Effelsberg) spectral index map.
\textit{Bottom left:} Spectral index from CGPS and Effelsberg data (1420 MHz$-$2725 MHz).
\textit{Bottom right:} Spectral index map combining all frequencies, as presented in Fig. \ref{fit_results_all}.
}
\label{all_spxs}
\end{figure*}

The best-fit radio surface brightness at 150 MHz ($S_{150}$, Fig. \ref{fit_results_all} upper left corner) is lower than that in our 
LOFAR data. This is of course because the integrated surface brightness at 143 MHz is substantially higher than predicted from the
higher frequency data. The spectrum has curvature and so cannot be accurately fit by a power law; we discuss this curvature
thoroughly in later sections of this paper.

In fact, the spectral index of an H II region also changes as it goes from being optically thin to optically thick. This could presumably
happen at LOFAR frequencies and so, in principle, there is no reason why a power law should describe the spectrum of OA 184 in the 
frequencies of relevance for this paper.

For this reason we made multiple spectral index maps using different combinations of the LOFAR, CGPS 408 MHz, CGPS 1420 MHz,
and Effelsberg 2695 MHz maps. These are shown in Fig. \ref{all_spxs}.

\section{OA 184}
\label{sec_oa184}

Although in earlier works OA 184 has been considered to be a SNR, \cite{foster06} presented compelling evidence that it is, in fact, an H II region, and OA 184 was
subsequently removed from the Green SNR Catalogue \cite[]{green17}. \cite{foster06} cited the following reasons for OA 184 not being a SNR:
\begin{itemize} 
\item It shows no X-ray emission.\item  It is not seen in low-frequency surveys \cite[e.g.][]{vessey98} because of a flat radio spectrum
($\alpha = 0.2$ to $\alpha = 0.14$).
\item It presents in IR surveys as a shell (see e.g. Fig. \ref{wise_rgb}).
\item In the optical spectrum H$\alpha$ dominates over other metal line emission \cite[]{fesen85}.
\item There is no polarised continuum emission at 21\,cm  \cite[]{kothes06}.
\item The radio recombination hydrogen lines are at a level consistent with a thermal shell \cite[]{foster06}.
\item There is a single O7.5V star within the shell with
a similar excitation parameter as the nebula \cite[]{foster06}.
\end{itemize}

The lack of X-ray emission could be consistent with a scenario
in which OA184 is a SNR that is too old to emit X-rays.
However, the lack of forbidden line emission from metals, the bright, clumpy infrared emission, and the recombination lines 
make it difficult to argue against the case that OA 184 is an H II region, likely energised by its central
star. 

\begin{figure}
   \centering
   \includegraphics[width=\hsize]{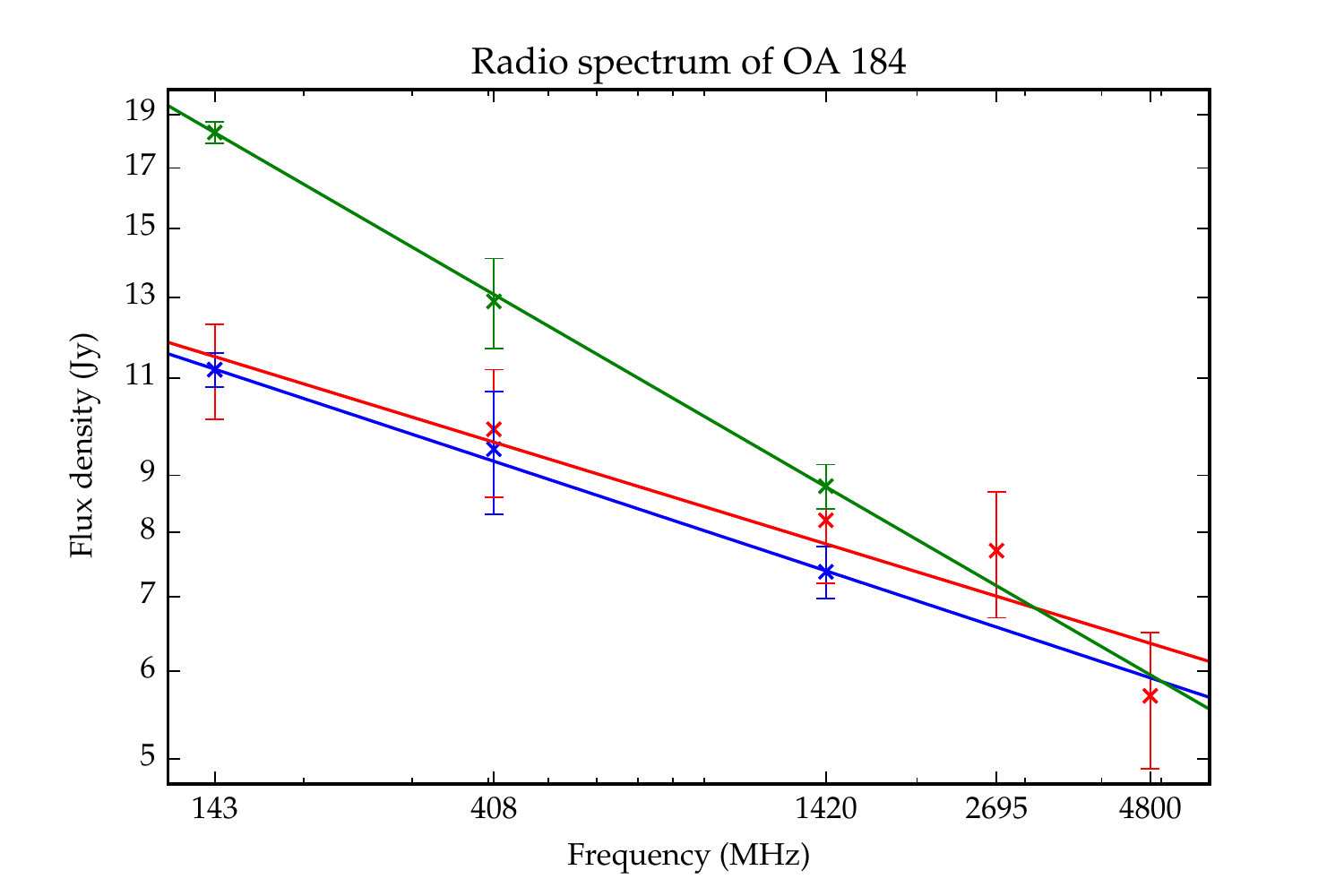}
      \caption{Flux density values of OA 184 including compact sources (green),
with the compact sources subtracted as described in \cite{foster06} (red), and with the compact sources subtracted as extrapolated
from the total flux density in compact components as measured with LOFAR (blue).
              }
         \label{oa184_spectrum}
   \end{figure}

Our LOFAR observations agree very well with the flat 
(thermal) radio spectrum derived by \cite{foster06}.
As mentioned by \cite{leahy05} 
and \cite{foster06} much of the integrated flux density measurement difficulty has to do with subtracting the emission from
compact sources in the face of the shell. We subtracted more sources from the total flux density of the OA 184 outline than
the earlier works because these typical sources have a spectrum $\alpha\sim0.7$ and therefore are more prominent in the LOFAR image.
We accounted for these differences in the following way: in our LOFAR image, we measured 7.1 Jy in compact sources. For a spectral
index $\alpha = 0.7$ we expect these to contribute 3.4 Jy at 408 MHz and 1.4 Jy at 1420 MHz. We subtracted these values to the total 
(OA 184 + compact sources) flux density values quoted in \cite{leahy05}. The fits for the flux density values of OA 184 with compact sources,
with the compact sources subtracted as described in \cite{foster06}, and with the compact sources subtracted as described above are
plotted in Fig. \ref{oa184_spectrum}.

\subsection{Structures within OA 184}

   \begin{figure}
   \centering
   \includegraphics[width=\hsize]{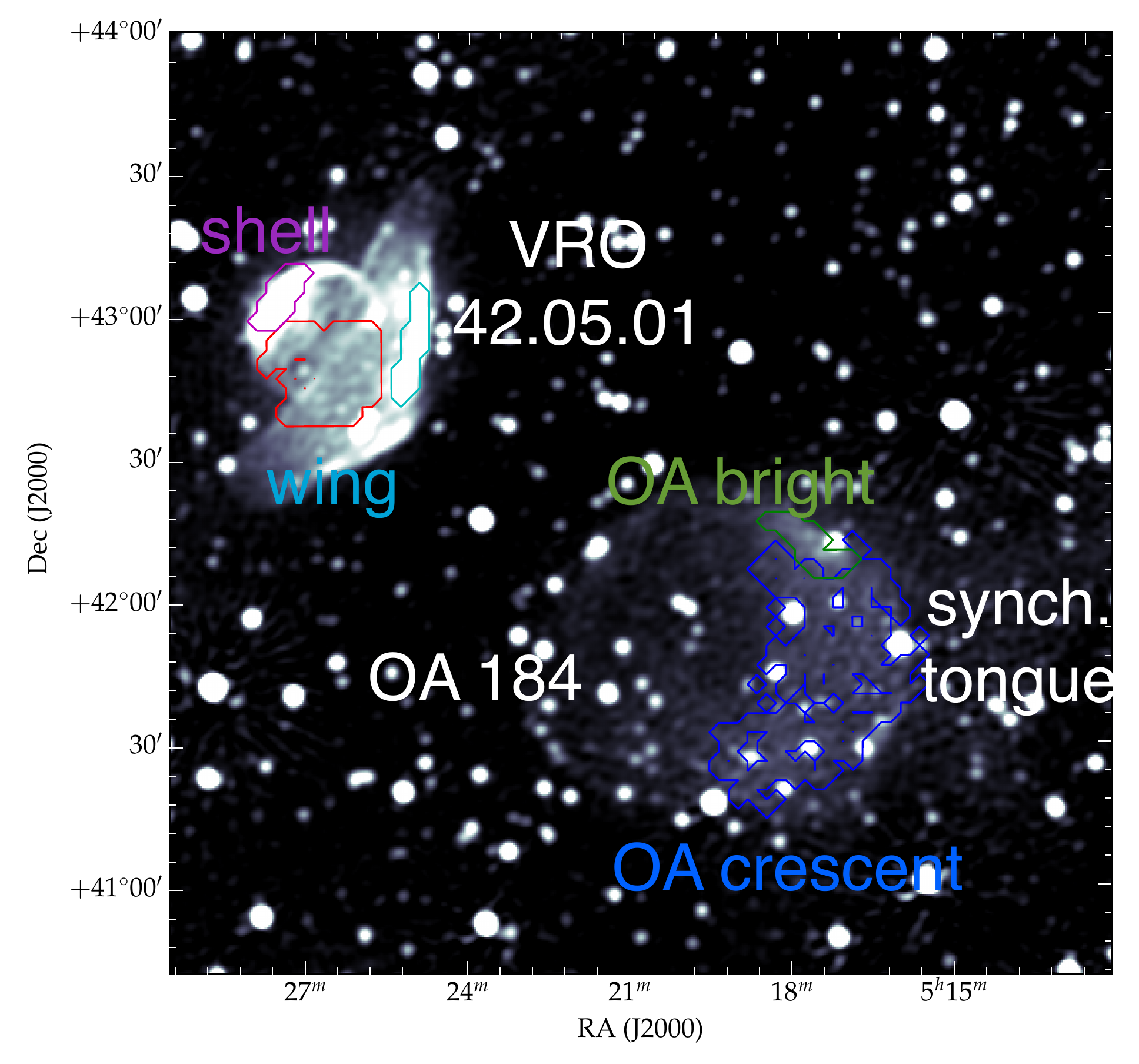} \\
   \includegraphics[width=\hsize]{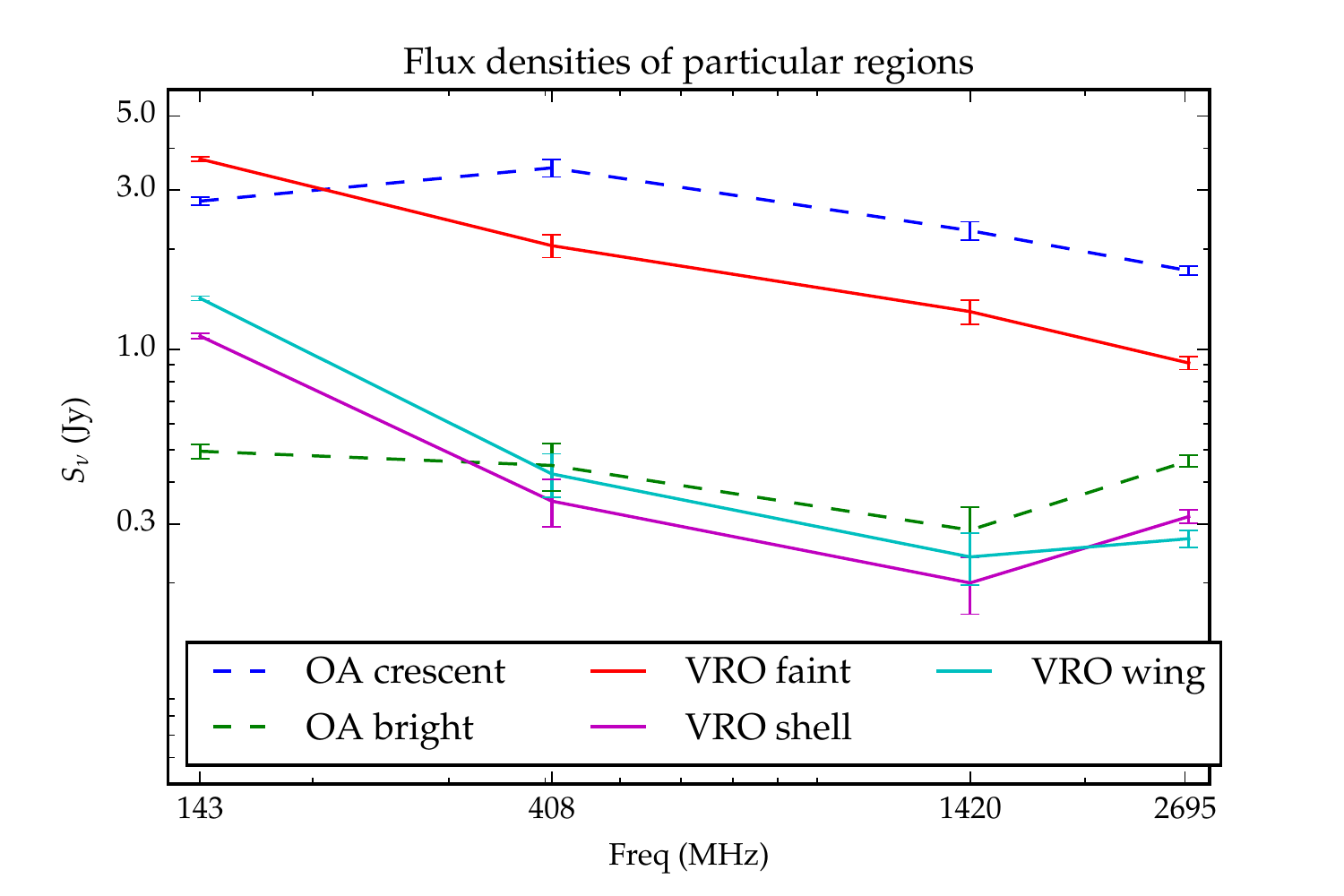}
      \caption{Flux density values of some regions in OA 184 and VRO 42.05.01.
      The colours in the plot correspond to the regions of the same colour in the map above.
              }
         \label{regions}
   \end{figure}

An aspect worth noting regarding the morphology of this source is that both the radio intensity (Fig. \ref{full_field})
and the spectral index behaviour at all frequencies (Fig. \ref{all_spxs}) suggest that there are three distinct regions in OA 184:
a faint region to the east, a brighter, crescent-moon shape
in the west, and a small bright spot in the north. 
This crescent moon-shaped feature is also responsible for the bulk of infrared emission from OA 184 
(see Fig. \ref{wise_rgb}), and has substantial morphological
correspondence with the neutral hydrogen emission seen at velocities of $\sim 27-36$\,km\,s$^{-1}$.

Fig. \ref{regions} shows the behaviour with frequency of the crescent-moon shaped region and a bright 
spot in the north of the source. In the case of the crescent, the H II emission becomes optically thick at around
400 MHz, whereas the bright region becomes optically thick at a lower frequency between the 143 MHz and 408 MHz
points. This can explain the diversity in spectral index behaviour for these regions in Fig. \ref{all_spxs}.

\subsection{The  synchrotron tongue}

\cite{foster06} proposed that the "tongue", the extended trail west of OA184, is a synchrotron filament that lies behind the H II region
shell. They saw polarised 6 cm emission both inside OA184 and to its west, where the  tongue is located, and there is a gap coincident with the edge of the
shell. They argued that the shell depolarises the background polarised emission from the  tongue. 
At the rim of the shell, where it is the thickest, is where the majority of the depolarisation occurs. They invoke a shell structure of thickness
$\sim 0.11\degr$ to account for the necessary rotation measure. 

Our LOFAR observations show no limb brightening,
which is expected if OA 184 were indeed such a shell. 
In fact, from our observations, it seems that the  tongue does not emit synchrotron at all.

In most of the spectral index maps in Fig. \ref{all_spxs} one cannot see the  tongue from the background. 
However, in the spectral index map including all the data (Fig. \ref{fit_results_all}), the  tongue has a spectral
index behaviour similar to the brighter, crescent-shaped region of OA 184.
In comparing our Fig. \ref{full_field}
to the full resolution
1420 MHz map in Fig. 1 of \cite{leahy05}, we can appreciate that the tongue is not any more prominent at 143 MHz than at 1420 MHz.
In fact, in addition to its flat spectral index, the tongue west of OA 184 displays a substantial amount of
infrared emission (see Fig. \ref{wise_rgb}), suggesting a thermal origin. This component could be linked to the structure
that appears in H I at $-25$ to $-33$ km\,s$^{-1}$ (see Fig. \ref{vel_plots}). For these reasons we
suggest that the  tongue might be a H II region.

The similarities in spectral index behaviour and infrared morphologies, as well as the fact that both structures
appear and fade at similar velocity ranges, could suggest that OA 184 and the  tongue are in fact part of the same structure;
for instance, the  tongue could be a breakout of the H II region or could have that shape owing to some projection effect.
However, \cite{foster06} found different polarisation properties in both structures, which complicates the suggestion that
they might be a single H II region.

\section{Analysis of VRO 42.05.01}
\label{spx_analysis}

\subsection{Comparisons with earlier work}

   \begin{figure}
   \centering
   \includegraphics[width=\hsize]{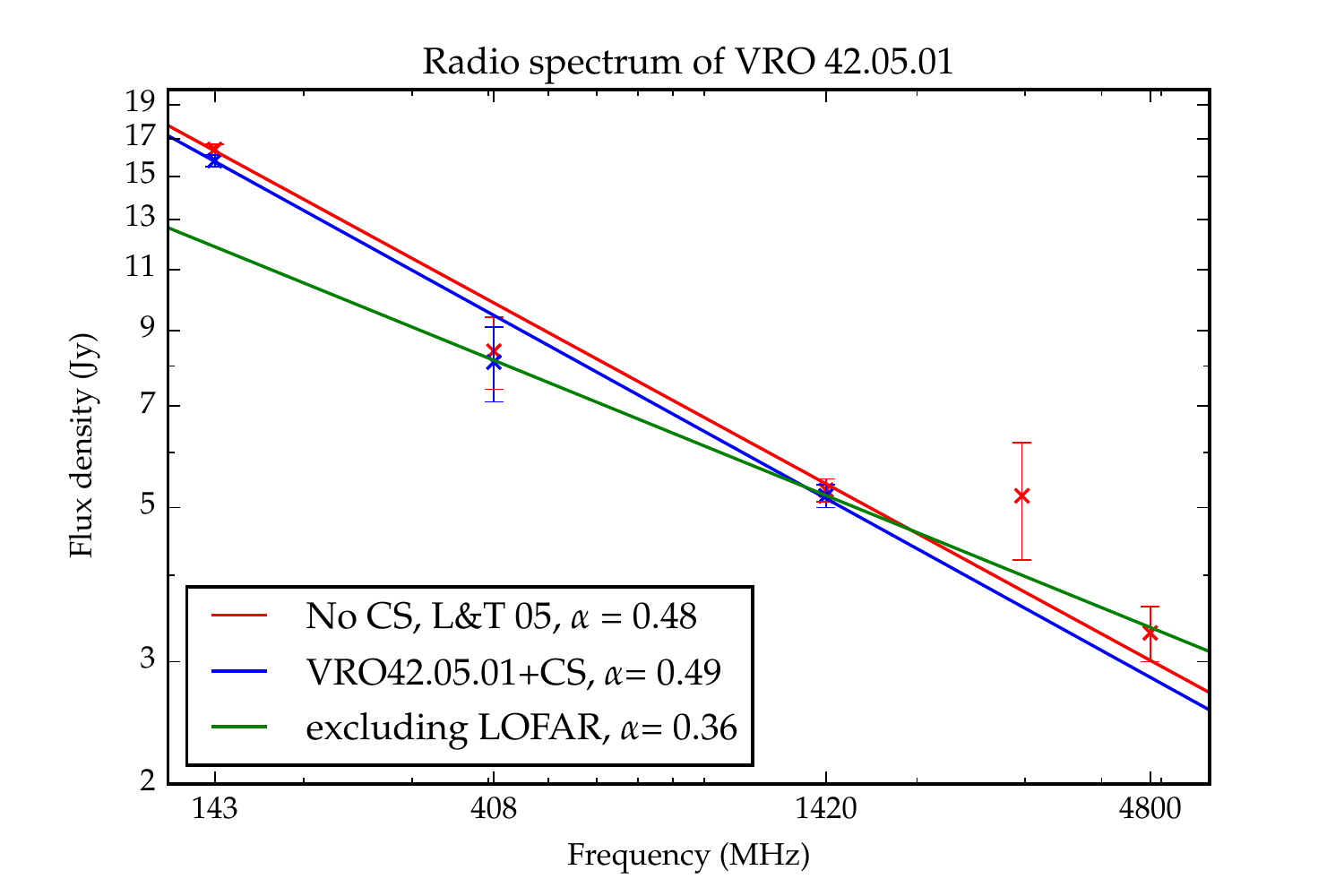}
      \caption{Flux density values of VRO 42.05.01 including compact sources (blue),
with the compact sources subtracted as described in \cite{leahy05} (red), and excluding the 143 MHz LOFAR data point (green).
The point at 4800 MHz is taken from \cite{gao11} and does not have the compact sources subtracted.
              }
         \label{vro_spectrum}
   \end{figure}

The LOFAR observations show that VRO 42.05.01 is significantly brighter at 143 MHz than the earlier spectral index calculations predict
\cite[see Fig. \ref{vro_spectrum}, plus spectral index calculations in][]{leahy05, gao11}. Even accounting for the presence of compact sources in the face of the remnant it is 
clear that the radio spectrum of VRO 42.05.01 steepens at low frequencies. We present possible interpretations for the curved spectrum of the source later in this paper (section \ref{sec_interpretation}).

\cite{leahy05} used the 1420 MHz and 408 CGPS data to conduct a thorough study  of 
the radio spectrum of VRO 42.05.01 and OA 184. These authors found that for VRO 42.05.01 
the shell and wing regions have different spectral indices: $\alpha_{\mathrm{shell}} = 0.31 \pm 0.03$ 
and $\alpha_{\mathrm{wing}}= 0.47 \pm 0.03$. This difference is noticeable in Fig. \ref{all_spxs}, bottom left.
The LOFAR-Effelsberg map (Fig. \ref{all_spxs}, bottom right), which has the largest
leverage arm in frequency, gives $\alpha_{\mathrm{shell}} = 0.41 \pm 0.02$ and $\alpha_{\mathrm{wing}} = 0.48 \pm 0.03;$
 these are slightly larger indices than the
\cite{leahy05} values, but preserve the difference between shell and wing.

We suspect that the difference in spectral index behaviour between the shell and the wing is in fact due to
a difference between faint and bright components. The plot in Fig. \ref{regions}
shows the flux densities at the frequencies of interest of three regions in VRO 42.05.01: the two very bright
regions in the north of the shell and the west of the wing, and a diffuse, faint component that extends between
both. The two brightest regions in the shell and the wing have very similar behaviour with a flat spectral
index at higher frequencies and an abrupt steepening at LOFAR frequencies. The faint component has a
roughly constant power-law shape with slope $\alpha_{\mathrm{faint}} = 0.48\pm0.02$.

It is possible that the faint region has the same steepening as the bright regions at LOFAR frequencies
but that it is not visible 
because of free-free absorption. This would require there being ionised gas covering
the surface area of only the red region  in Fig. \ref{regions}, and not the cyan nor the magenta;
this is unlikely 
since the X-ray papers \cite[]{burrows94,guo97} have found a very low column density towards VRO 42.05.01.
Moreover, for free-free absorption to be significant at 150 MHz, 
the purported ionised gas would have to have a high emission measure, which would likely
result in radio, optical, or infrared emission.
We believe that the difference in spectral behaviour of steep and faint components is intrinsic and due to the
shock physics at play, which we discuss in section \ref{sec_interpretation}.

\subsection{Relation between radio spectral index, radio brightness, and H$\alpha$ brightness}

   \begin{figure}
   \centering
   \includegraphics[width=\hsize]{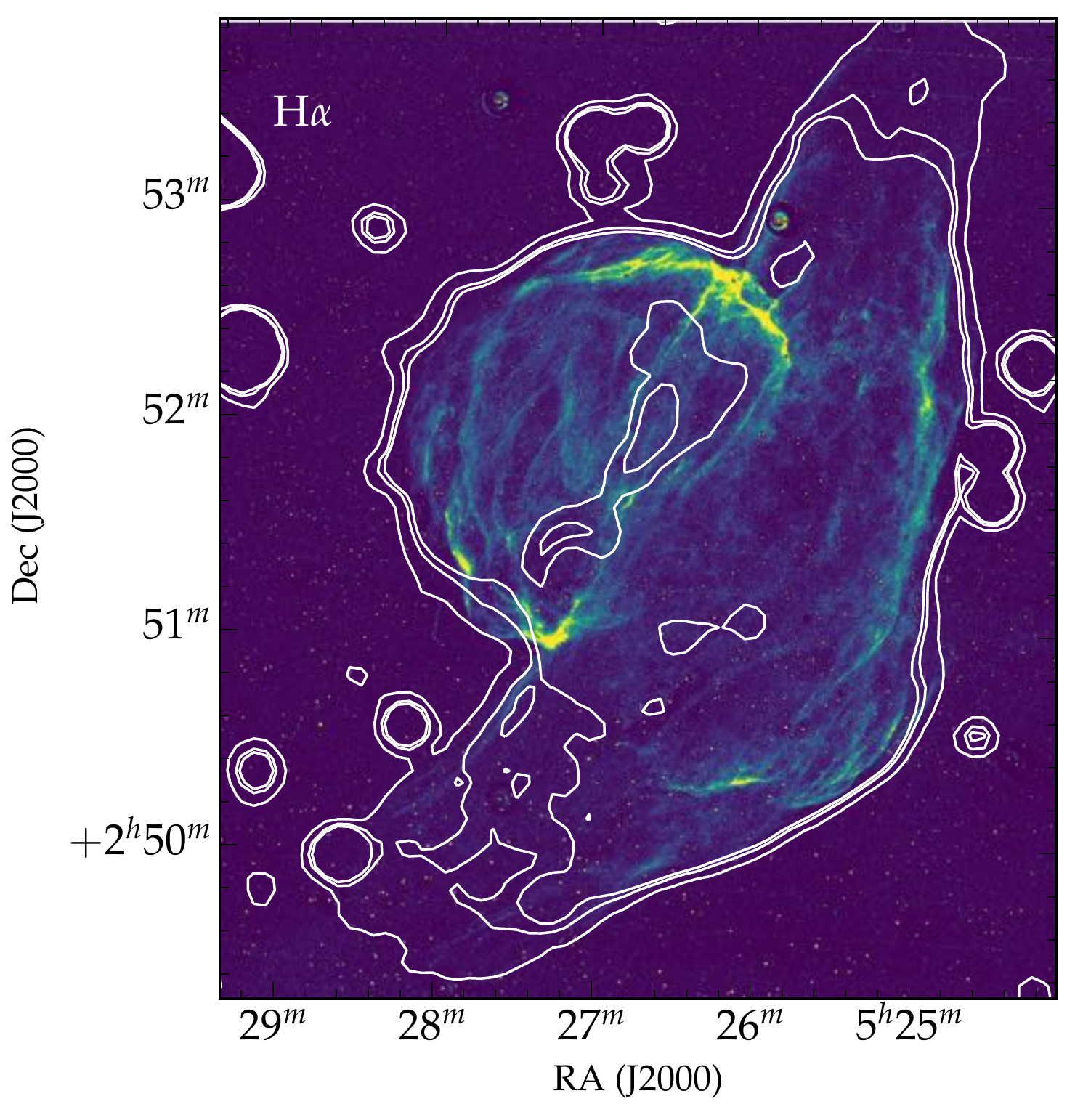}
      \caption{H$\alpha$ map overlaid with 143 MHz contours. 
      Taken from the Middlebury Emission-Line Atlas of Galactic SNRs (Winkler et al., in preparation).
              }
         \label{optical}
   \end{figure}
   
      \begin{figure}{
   \centering
            \includegraphics[width=\columnwidth]{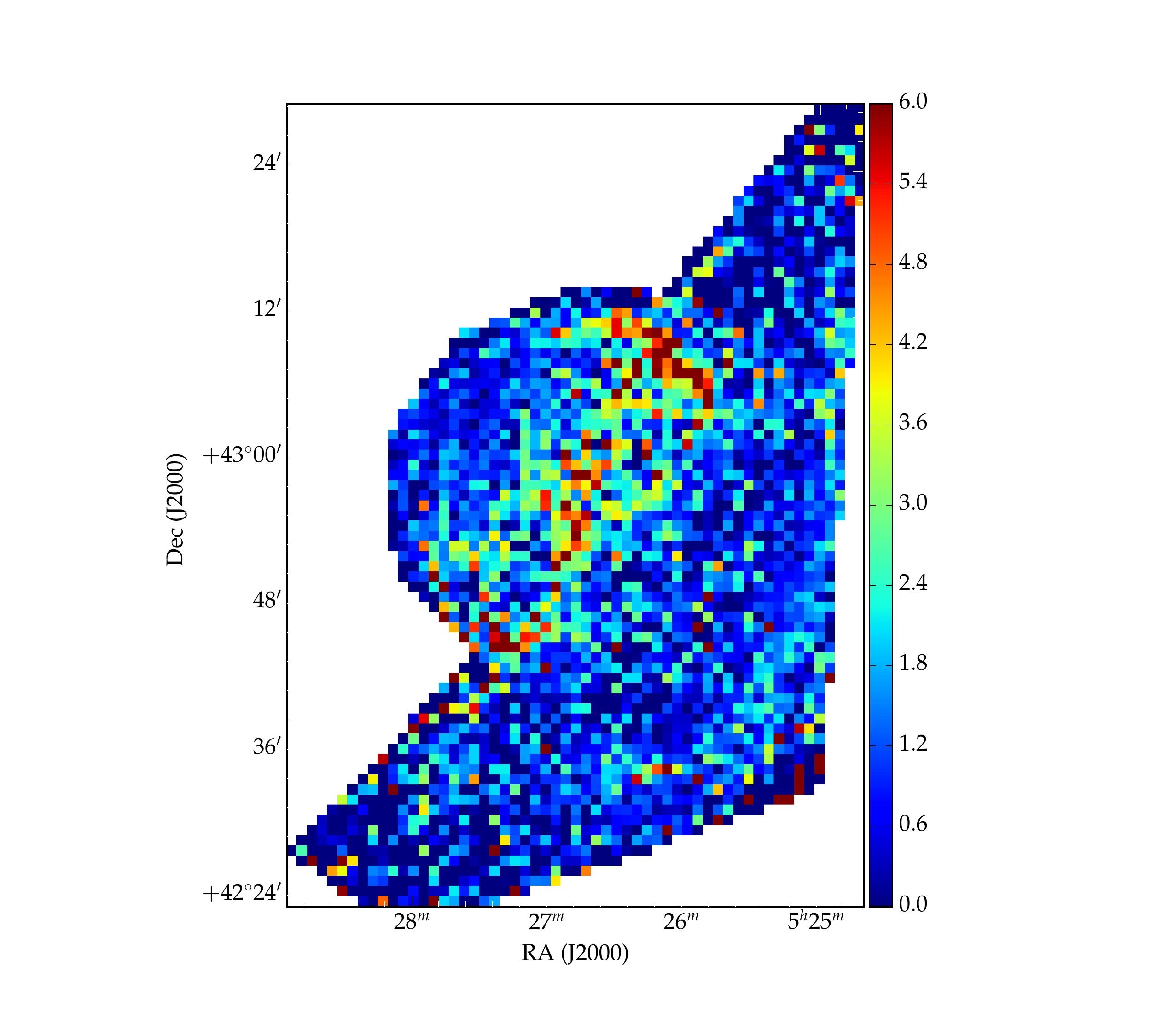}
      \caption{
              Ratio of optical to radio emission. The H$\alpha$ map in Fig. \ref{optical}
was resolved down to match the radio image, and the emission within the masked region of VRO 42.05.01 was normalised
to one for both the optical and radio maps before taking the ratio.
}
         \label{ratios}}
   \end{figure}

In the case of radiative SNRs, optical emission traces low-density atomic gas that is rapidly cooling. 
Fig. \ref{optical} shows the morphology of VRO 42.05.01 in H$\alpha$ emission. It is striking 
how much the optical morphology resembles the radio morphology, outlining both the wing and the shell.  
Fig. \ref{ratios} further emphasises this point, showing a roughly constant ratio of optical to radio emission,
save for a region of excess optical emission in the line where the shell and the wing meet.

In general, the coincidence of the synchrotron-enhanced radio emission with very strong optical
filaments in lines suggests that the optical features delineate the position of cooling post-shock ISM gas.
The region of excess optical emission along with the presence of the two very bright optical knots
in the junction where the two structures cut suggests that we are looking at a circular rim that is perpendicular
to the plane of the sky and has swept a substantial amount of hydrogen but  is rather thin (and
so emits little synchrotron radiation relative to the rest of the remnant). 
It is possible to think of a sphere (the shell) and a cone (the wing) that meet in this optically emitting rim that is
a boundary between both structures and the ISM.
If this is the case, then the optical and radio emission in the wider side of the wing require that
the base of the cone has a sharp edge, which is difficult to form if the wing is the result of a shock breaking into a 
region of lower density, as suggested by 
\cite{pineault85}. 

The region of excess optical emission, moreover, appears to have a flatter spectral index than the
rest of the SNR in all the maps in Fig. \ref{all_spxs}.
Although from the optical map the emission in the boundary between the shell and the wing
appears to be a limb brightening effect, it is not clear why the limb brightening
would not affect the radio emission. There must be some additional physical effect at play that enhances the optical 
emission or suppresses and/or flattens the synchrotron emission.

\subsection{Environment and interaction}
\label{env_inter}

      \begin{figure*}
      \vspace{-0.3cm}
      \centering
            {\includegraphics[width=0.87\textwidth]{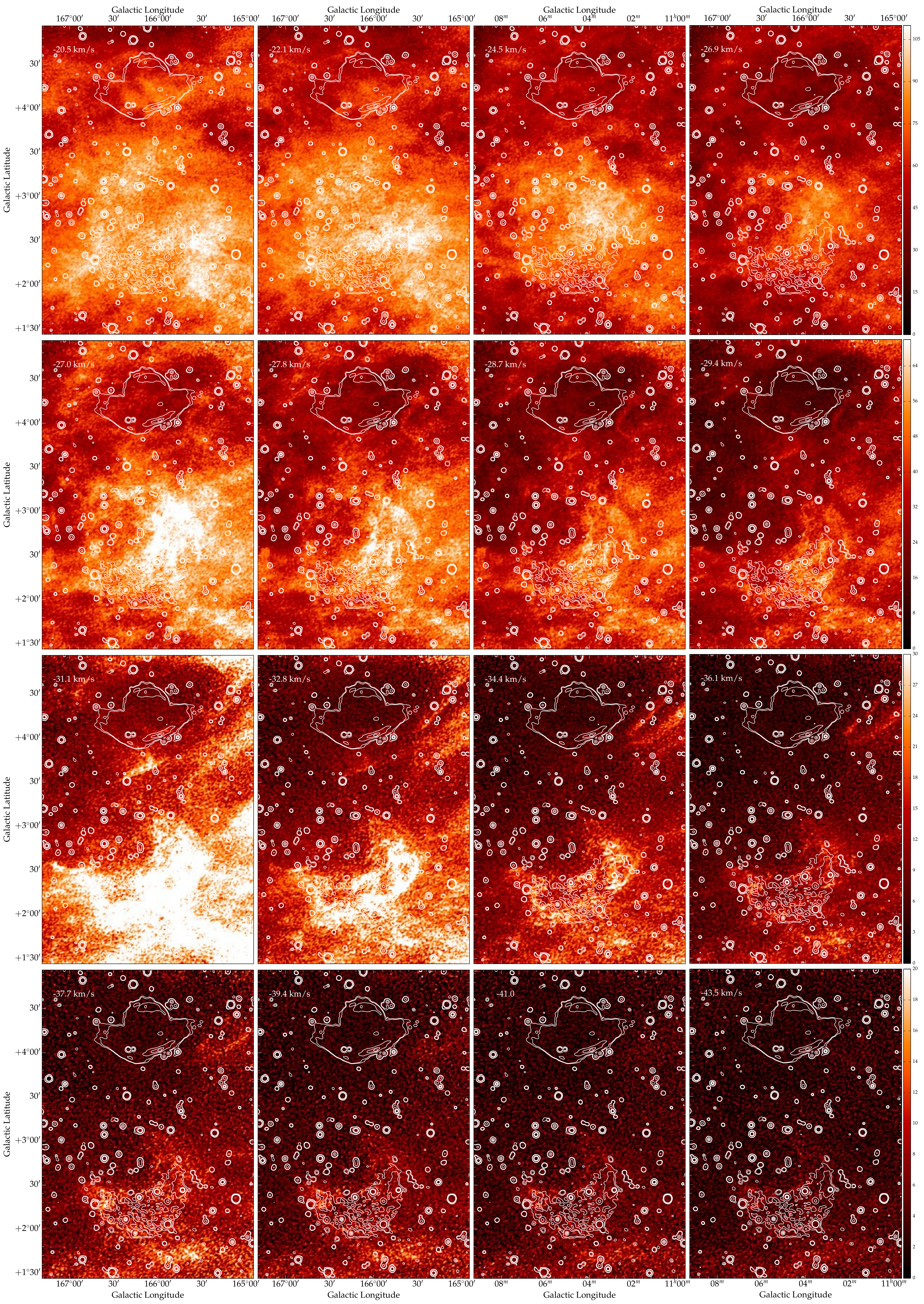}
      \caption{Velocity distribution of HI emission towards VRO 42.05.01 and OA 184. 
      The data cube was taken from the CGPS. The radio contours at 143 MHz from the LOFAR
      map in this work are overlaid in white.
      The wedges in the upper left corner of each plot correspond to the velocity of each slice. 
      The brightness temperature $T_B$ colour scale is the same for each row and goes as indicated by the colour bar to the left of each row, but it
      is not the same between rows. The ranges are as follows: first row: $0-110$\,K; second
      row: $0-70$\,K; third row: $0-30$\,K; and fourth row: $0-20$\,K. We have saturated the emission from OA  184 (bottom) to highlight the
      fainter emission in the surroundings of VRO 42.05.01.
              }
               \label{vel_plots}}    
   \end{figure*}

There are several important hints suggesting that VRO 42.05.01 might be evolving within an environment of different densities.
 \cite{landecker82}, \cite{pineault85, pineault87}, and  \cite{landecker89} all invoked an inhomogeneous medium to explain  the
 unusual morphology of VRO 42.05.01. The latter paper analyses interferometer and single-dish data of 21 cm line emission in 
  the line of sight of VRO 42.05.01 and identifies multiple features at different velocities that seem to match the morphology of the source
 and that, according to the authors, establish interaction. The \suzaku\ observations in \cite{matsumura17} show that the X-ray spectrum of the shell side 
 of the remnant is described by a recombining plasma, and the authors argue that it is because in this side the remnant is in contact with cool, dense gas.
 They also point out that the emission  measure suggests higher ambient gas density in the shell region.
 \cite{leahy05}, who found that the shell has a lower spectral index than the wing,
 suggested that different Mach number values in each of the shock fronts 
 are responsible for this feature. In this case, the lower spectral index for the shell would be explained by a higher density surrounding it.
 
 There are also some reasons to be suspicious that these inhomogeneous density conditions do in fact describe the
 environment of VRO 42.05.01. One reason is that the surface brightness of VRO 42.05.01 is relatively low for an SNR
 evolving in a dense medium. Secondly, 
the $\gamma$-ray spectrum in \cite{araya13} is best explained by inverse Compton emission from relativistic
electrons, and not from the decay of neutral pions produced from accelerated cosmic rays interacting with a nearby molecular cloud.
Also, the \cite{landecker89} argument is based on morphological similarities of the H I and SNR features, 
which, although suggestive, are no physical proof of interaction. 
The CO survey of \cite{huang86} did not show signs of sharply differing densities around the remnant, nor
any clear molecular cloud. To our knowledge, there has been no study of molecular line spectra 
in the direction of the SNR for 
broadenings, wings, asymmetries, or ratios of different excitation states in the lines, which would constitute an unambiguous 
proof that the SNR is interacting with a nearby molecular cloud.

The CGPS included an atomic hydrogen survey made with the Dominion Radio Astronomical Observatory (DRAO) interferometer
complemented with data from the DRAO 26 m single dish. 
Incidentally, this is the same telescope set-up used in the \cite{landecker89} paper, whose data 
is sensitive to all structures of angular scales down to their 4\arcmin\ resolution.
By the time the CPGS data was taken, the interferometer had gone through several upgrades, resulting in a higher
angular resolution of 1.7\arcmin.

Having said that, much of the  \cite{landecker89} argument for interaction is based on features that are not
visible in the CGPS data. In Fig. \ref{vel_plots} we present the velocity maps for H I emission in the region of VRO 42.05.01 and OA 184.
The velocities go from $-20.5$ km\,s$^{-1}$ to $-43.5$ km\,s$^{-1}$. We plotted this range of velocities because \cite{landecker89} 
found that the systemic velocity of the SNR is  
$\sim -34$  km\,s$^{-1}$. \cite{landecker89} list a series of H I features, but some do not appear in the CGPS data:
\begin{itemize}
\item They locate a minimum of H I emission coincident with the position of VRO 42.05.01. at $-28$ km\,s$^{-1}$, which they identify 
with the interstellar cavity that the SNR has re-energised. In the CGPS data there is not a minimum at these velocities, 
rather the emission keeps decreasing at velocities more negative than $-28$ km\,s$^{-1}$.
\item They find that the cavity that most closely resembles the shape of the SNR is at $-36$ km\,s$^{-1}$. 
Although we see a hole of emission in the CGPS data at those velocities in the position of the SNR,
it is not evident that the shape of the cavity matches the shape of the SNR.
\item They see a semi-circular arc defined by the continuum shell that appears to be completed by an H I feature at $-38$ km\,s$^{-1}$.
\item They see a structure at -42 km\,s$^{-1}$ (\lq structure K' in their paper) that extends over the surface of the 
wing and that allegedly is accelerated gas associated with the cavity. 
\item They also find an expanding half shell from $-34$ to $-40$ km\,s$^{-1}$ (\lq structure J'), which they attribute to the rings formed as the shell
is cut along the velocity axis.
\end{itemize}

The CGPS images are noise-limited with an rms brightness temperature $\Delta T_{B} \sim 3$K in each 0.82 km\,s$^{-1}$ channel \cite[]{taylor03},
and so in principle should be able to detect emission from the features described in \cite{landecker89}. 
In addition to these CGPS data not being clearly indicative of 
interaction, there do not appear to be two different neutral hydrogen density conditions that cut across the remnant at its alleged systemic velocity
parallel to the Galactic plane.

The remnant, however, is at the boundary of a large polarised bubble visible in the CGPS polarisation data \cite[]{kothes04}.
The large structure, with an angular size of 9\degr, is clearly visible in the Stokes U and Q images, but only barely in
polarised intensity and has no counterpart in Stokes I, which means it is likely a Faraday screen. The breakout boundary
of the SNR (the line between the shell and the wing) coincides with the edge of the polarisation bubble. The authors
propose that it might be the remains of an old SNR whose highly compressed magnetic field still affects the
background emission. This would be a way of having an environment with two sharply different densities.

In summary, many models that explain observed properties of VRO 42.05.01 are based on the SNR expanding into and interacting
with media of different densities. However, most of the arguments in favour of association are based on morphological similarities
and positional coincidence, and
there is still no conclusive physical proof that the conditions around the remnant are 
in fact so inhomogeneous.

\section{Interpretation}
\label{sec_interpretation}

In the interpretation we discuss some possibilities for the spectral shape and 
spatial morphology of
VRO 42.05.01.

\subsection{Possible reasons why VRO 42.05.01 has a curved spectrum at low frequencies}
In this section, we elaborate on some possible causes for the curved spectrum of VRO 42.05.01.
In fact, many MM SNRs have radio spectral indices that are flatter than the canonical $\alpha = 0.5$ predicted by standard shock acceleration theory \cite[]{bell87}, 
and there is still no satisfying explanation for the existence of values of $\alpha < 0.5$. 
Therefore, to explain the radio spectrum of VRO 42.05.01 we need a mechanism that 
is responsible for a flat spectrum with low-frequency steepening. We explore the following scenarios:
high compression ratio shocks, two electron populations with different spectral indices that dominate at different frequencies, 
two shock regimes,
and re-acceleration of previously accelerated electrons. 
We favour the high compression ratio shocks scenario, as it explains both how the flat spectrum originates and
its steepening at low frequencies. However, we only briefly sketch it as a possible cause for the observed properties
of VRO 42.05.01, and we caution that detailed simulations are required for it to be a convincing explanation of
the presence of flat spectral indices in MM remnants.

\subsubsection{High compression ratio shocks}

The shock compression ratio $\chi$ is the ratio of postshock to preshock densities. 
For linear diffusive shock acceleration (DSA), the electron energy index $p$ is independent of energy 
and depends only on the shock compression ratio.
For an electron distribution with $N(E) dE = \kappa E^{-p} dE$, where the radio spectral index is $\alpha = \frac{p-1}{2}$, 
test-particle DSA predicts $p = \frac{\chi+2}{\chi -1}$ \cite[]{bell78a}. A strong shock has $\chi =4$,
corresponding to $\alpha=0.5$.

Radiative shocks can give rise to high compression ratios, as the surrounding medium compresses the plasma while it loses energy. 
For isothermal shocks, compression ratios go as the square of the Mach number \cite[$\chi = \gamma M^2$,][]{draine11}, 
so in principle very high compression ratios can be achieved even for modest Mach numbers, although in these 
cases the compressed magnetic field is likely to provide pressure support. Radiative shocks can be seen in the optical as bright, 
thin filaments. In many SNRs only parts of the shells are radiative, which may be the result of a non-uniform medium;
the portions of the 
remnant expanding into a denser medium enter the radiative phase earlier and are thus visible in the optical \cite[]{blondin98}. 

In these environments, the high $\chi$ does not happen immediately at the shock, but in an extended region behind it \cite[]{raymond79}. 
Electrons of different energies are scattered further within the region of increasing $\chi$, and hence feel higher effective compression ratios. 
This is somewhat analogous to the mechanism of non-linear shock acceleration upstream of the shock, only
this case affects the downstream region.
The idea behind this mechanism as an explanation for the low-frequency steepening of the flat radio spectrum of VRO 42.05.01 is that as electrons of varying energies scatter back and forth along
the shock front, they sample different compression ratios and therefore have a different value of the spectral index. 
The low-frequency electrons scatter closer to the shock front, experience a lower compression ratio, 
and so have a correspondingly higher index.

The diffusion length scale for an electron accelerated to very high energies is
\begin{equation}
l_\mathrm{diff} = \frac{D}{v} =  \frac{1}{3} \, \eta \, \frac{E c}{e B} \, \frac{\chi}{v_\mathrm{sh}},
\end{equation}
where $D$ is the diffusion coefficient, $v = \frac{v_\mathrm{sh}}{\chi}$ is the velocity upstream of the shock, 
$\eta$ is a parametrisation factor ($\eta=1$ corresponds to Bohm diffusion),
$E$ is the energy of the electrons, $c$ is the speed of light, $e$ is the elementary charge, and
$B$ is the magnetic field. For synchrotron electrons, the critical emitting frequency is 
$\nu_\mathrm{c} = 1.8 \times 10^{18} E^2 B$ \cite[]{ginzburg65}, and so we can expect that at 150 MHz the electrons emitting synchrotron
radiation in VRO 42.05.01 have a diffusion length-scale of
\begin{equation}
l_\mathrm{diff} = 1.2 \times 10^{15} \, \eta \, \left( \frac{B}{10 \, \mu\mathrm{G}}\right)^{-1} \left(\frac{\chi}{4}\right) \, \left( \frac{v_\mathrm{sh}}{200 \, \mathrm{km\,s}^{-1}} \right)^{-1} \mathrm{cm},
\end{equation}
and at 1500 MHz they have instead 
\begin{equation}
l_\mathrm{diff} = 3.8 \times 10^{15} \, \eta \, \left( \frac{B}{10 \, \mu\mathrm{G}}\right)^{-1} \left(\frac{\chi}{4}\right) \, \left( \frac{v_\mathrm{sh}}{200 \, \mathrm{km\,s}^{-1}} \right)^{-1} \mathrm{cm}.
\end{equation}

In the models of \cite{raymond79}, for cooling shocks in the interstellar medium, the temperature conditions in the shock vary
roughly an order or magnitude between the length scales of $1\times 10^{15}$ cm and $4\times 10^{15}$ cm \cite[see e.g. Figs 1 and 2 in][]{raymond79}.
This could account for the difference in spectral index at 150 MHz and 1500 MHz.

Of course, this is a simple order of magnitude estimate. The \cite{raymond79} models assumed an ambient density of $n = 10$ cm$^{-3}$,
which seems rather high for VRO 42.05.01, and we do not know what is an appropriate value of $\eta$ ($\eta \lesssim 10$ would correspond
to a very turbulent medium). 
In practice, non-linear effects can also be important \cite[]{ellison85}.
\cite{onic13} linked high compression rations with flat spectra for radiative shocks.
However, the situation of high compression ratios downstream of the shock 
has not been,
to our knowledge, exhaustively dealt with
in the theoretical literature, unlike the upstream case.
We simply present this scenario as a possible explanation for the observed behaviour of the radio spectrum of  VRO 42.05.01, and we note that further analysis is required both from theoretical and observational standpoints.

\subsubsection{Different electron populations and different shock regimes}

A natural explanation for variations in the radio spectral index is that there are two populations of emitting electrons,
one with a steeper index and one with a flatter index, and each dominates at different frequency regimes.
We find in section \ref{spx_analysis} that the region with fainter radio emission has a 
roughly constant spectral index at low radio frequencies, and that
the brighter regions steepen abruptly between 408 MHz and 143 MHz.
If this is due to two electron populations, the steep index population that becomes more dominant at low frequencies
would need to be confined within the brightest regions. 

The two electron populations can be due to different shock regimes that accelerate electrons
with different energy indices and that are both present in the remnant. This would require that, in addition to the
 radiative shocks  that outline the entire shell as evidenced from optical emission,  there
 is another type of shock in the brightest regions.
Another circumstance that can give rise to two different electron populations is that ISM electrons are compressed 
and accelerated \cite[i.e. the van der Laan mechanism,][]{vanderlaan62}. In this way, both the shock-accelerated
and the compressed electrons are present, and each might have an intrinsically different spectral index. \cite{rogers08}
proposed that interstellar electrons have $\alpha = 0.5 \pm0.1$ between 100 and 200 MHz, which would mean that interstellar electrons contribute 
the steeper spectral component and SNR electrons contribute the flatter spectral component ( Fig. \ref{vro_spectrum} shows that
$\alpha = 0.37$ excluding the data point at 143 MHz).

The presence of two electron populations can account for the steepening of the spectrum, but we still 
lack an explanation as to how the population with a spectral index 
$\alpha < 0.5$ arises.

\subsubsection{Re-acceleration}
Another method for creating a flat-spectrum electron population is the re-acceleration scenario proposed in
\cite{uchiyama10}.
They showed that in MM SNRs interacting with molecular clouds, the re-acceleration of pre-existing cosmic-rays by DSA at a cloud shock is 
sufficient to power the observed $\gamma$-ray emission through the decay of neutral pions. 
These authors proposed that the radio emission may be enhanced by the presence of secondary electrons and positrons, 
that is, the products left over from the decay of charged pions. 
According to this model, the synchrotron radiation from the secondary electrons contributes most of the emission at lower radio frequencies, 
whereas that from the primary electrons (which naturally have a steeper spectrum) contributes most of the emission at higher radio frequencies. 
This model is at odds with our observations, since it predicts a flat, rather than a steep, spectrum at LOFAR
frequencies.


\subsection{Is there a bow shock in VRO 42.05.01?}
\label{bow_shock_section}

The morphology of VRO 42.05.01 is extremely puzzling, and none of the proposed scenarios explain all of the
observed properties of the source. In particular, the shape of the wing is vexing --it consists of almost straight lines forming an isosceles
triangle with a vertex angle of 120$^\mathrm{o}$ with all sides emitting
in the optical and radio, requiring sharp boundaries in all directions. The models in the \cite{landecker82} and \cite{pineault85, pineault87} papers 
suggest that the explosion happened in the denser environment of the shell and eventually part of the shock broke
out to a lower density region, forming the wing. This scenario cannot explain that the wider side of the wing emits strongly in optical 
and radio.

In this rather speculative section we present a modification of this scenario that could explain the sharp outline of the wing and
the fact that it dissipates in the base angles.

Consider a massive star that is moving supersonically in a dense ISM region and that crosses the
boundary to a more diffuse region perpendicularly, creating a small hole in the boundary separating the two regions.
This star causes a bow shock. If the star is moving at 1.15 times the ISM sound speed, it forms a Mach cone of angle 120$^\mathrm{o}$. 

      \begin{figure*}
      \centering
            {\includegraphics[width=0.9\textwidth]{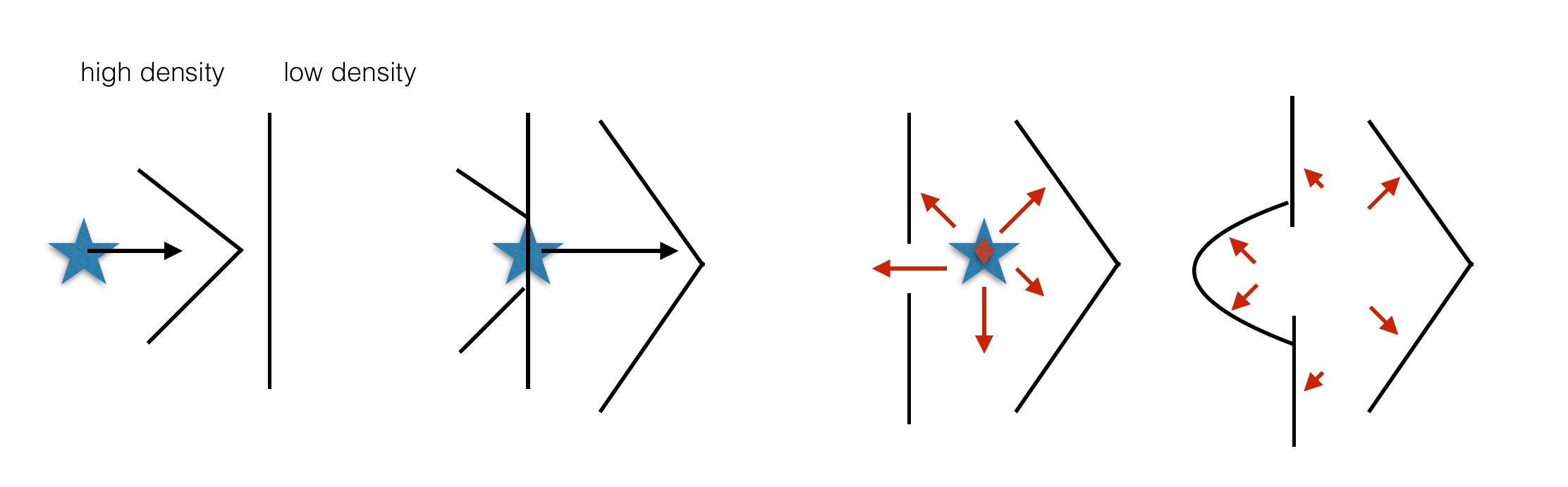}
      \caption{Cartoon depicting our proposed scenario for the formation of VRO 42.05.01 as discussed in section \ref{bow_shock_section}.
      The progenitor of VRO 42.05.01 would have been travelling between media of two different densities, 
      at a speed approximately 1.15 times the local sound speed in the ISM in the low-density region, creating 
      a Mach cone of 120$^\mathrm{o}$. When the supernova explosion occurred, the ejecta would have expanded
      in the triangular cavity caused by the bow shock, and through the small cavity to the high-density region.
              }
               \label{cartoon}}    
   \end{figure*}

If the star explodes shortly after crossing the boundary between the dense and diffuse regions, it will have exploded 
in a triangular cavity (the sides of the triangles are caused by the sides of the Mach cone, and the boundary between
the two density regions), forming the wing. The base of the triangle will have a cavity due to the passage of the star,
where the ejecta could expand into the higher density medium, more slowly than they do in the low-density triangle of the
wing. This would explain the spherical, small shape of the shell. We depict this sequence of events in a cartoon in Fig. \ref{cartoon}.


This scenario needs to be explored further with hydrodynamical simulations in order to be credible.
Moreover, it 
shares the main caveat of the 
\cite{landecker82} and \cite{pineault85, pineault87} model; namely that no conclusive evidence for a sharp density gradient
has been found, as discussed in section \ref{env_inter}. Targeted, sensitive molecular line observations 
towards the edges of this SNR could convincingly establish interaction between the remnant and a neighbouring
molecular cloud, or at least determine that there is in fact a molecular cloud in the environment of the remnant.
Barring these limitations, we think that this is a plausible modification to the traditional model for the shape of VRO 42.05.01 that can 
explain more of the observed features without any additional difficulty.

\section{Conclusions}

In this paper we present a LOFAR map of the region of the Galactic plane centred at $l=166 \degr, \,b=3.5$\degr at 143 MHz, with a 
resolution of 148$\arcsec$ and an rms noise of 4.4 mJy\,bm$^{-1}$. Our map is sensitive to
scales as large as 6$^{\mathrm{o}}$. We also make use of archival higher frequency radio images from the CGPS at
408 and 1420 MHz, $IRAS$ infrared observations, and optical data. We analyse the two extended sources in our field, 
SNR VRO 42.05.01 and the H II region OA 184. We conclude the following:
   \begin{enumerate}
      \item VRO 42.05.01 has a higher flux density at LOFAR HBA frequencies than expected from reported values of the
      radio spectral index, resulting in a pretty significant increase of its radio spectral index at low frequencies, from the $\alpha = 0.36 \pm 0.06$
      derived in \cite{leahy05} from observations at 408 MHz and 1420 MHz, to  $\alpha = 0.49 \pm 0.02$ when we include the LOFAR flux
      density measurement at 143 MHz.
      \item The flux density of OA 184 at LOFAR HBA frequencies agrees very well with the spectral index found by earlier
      works, further supporting the status of the source as an H II region. 
      \item The  synchrotron tongue to the west of
      OA 184 does not have a more apparent radio luminosity at 143 MHz than at 1420 MHz and matches structures in the infrared and neutral hydrogen
      maps.
      We suggest that it is not a synchrotron source, but rather thermal in nature.
      \item The faint diffuse emission in VRO 42.05.01 appears to have a constant spectral index of $\alpha = 0.48\pm0.2$ between 143 and 2695 MHz.
      The two very bright regions north and west of the source have a remarkably flat spectral index between 408 and 2695 MHz and steepen 
      dramatically at LOFAR frequencies.
      \item The radio and optical morphologies of VRO 42.05.01 are remarkably well matched, except for a region of excess optical emission
      in the interface of the shell and the wing. This region has overall a flatter spectral index than the rest of the SNR.
      \item We explore several possible reasons for the spectral index of VRO 42.05.01 to steepen at low radio frequencies. We favour an explanation 
      whereby the radiative shocks have high compression ratios, and electrons of different energies probe different length scales across the 
      shocks, therefore sampling regions of different compression ratios.
      \item We propose that VRO 42.05.01 is the remnant of a star that was moving supersonically in an inhomogeneous medium. The wing is explained
      as the expansion of the ejecta in the Mach cone formed by the progenitor star.
   \end{enumerate}
   
\noindent
VRO 42.05.01 is a mysterious source, and, 
as is often the case, our new observations leave many questions about its nature unanswered. 
Molecular observations are required in order to unambiguously establish physical interaction between the
remnant and its environment, and hydrodynamical simulations are key to understanding how the source
developed its peculiar shape.
Further observations of SNRs with 
the current generation of low-frequency radio interferometers will shed light on how unique 
a feature the observed low-frequency steepening  is to VRO 42.05.01 or to the 
mixed-morphology class. 
Exploring this regime is useful 
not only because it is where SNRs are the brightest, but also because both
the environment and the physical processes at play in the shock can have an imprint on the low-frequency
spectrum as departures from the synchrotron power law shape.

\begin{acknowledgements}
We thank the unnamed journal referee whose helpful and enegaging comments improved the paper.

The work of MA and JV is supported  by a grant from the Netherlands Research School for Astronomy (NOVA). GJW gratefully acknowledges support from The Leverhulme Trust.

This paper is based (in part) on data obtained with the International \lofar \, Telescope (ILT). \lofar \, \cite[]{vanhaarlem13} is the LOw Frequency ARray designed and constructed by ASTRON. It has facilities in several countries, which are owned by various parties (each with their own funding sources) and are collectively operated by the ILT foundation under a joint scientific policy. \lofar\ data reduction used the DRAGNET GPU cluster (at the CIT in Groningen), which was funded by the European Research Council under the European Union's Seventh Framework Programme (FP7/2007-2013) / ERC grant agreement nr. 337062 (PI: Hessels). The research presented in this paper has used data from the Canadian Galactic Plane Survey, a Canadian project with international partners, supported by the Natural Sciences and Engineering Research Council.
\end{acknowledgements}

\begin{appendix}\section{Narrow-band images}

We present as an Appendix the three narrow-band LOFAR HBA images. 
   \begin{figure*}
            {\includegraphics[width=\textwidth]{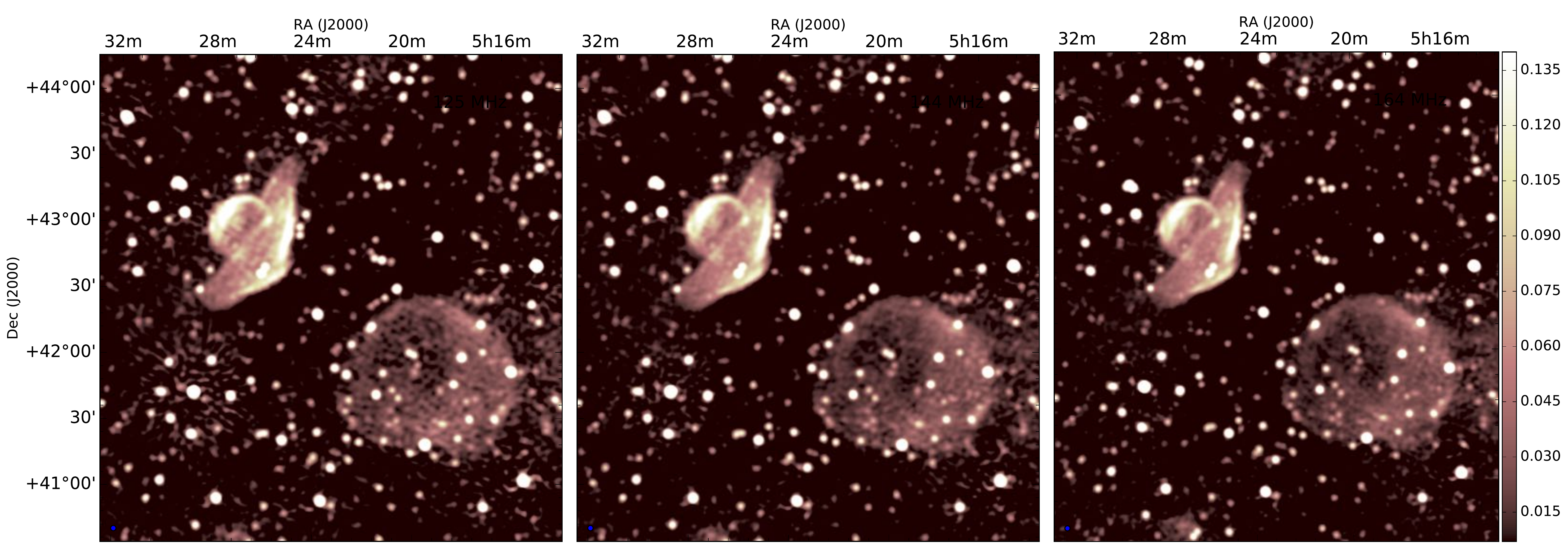}
      \caption{Uv-matched, narrow-band images at 125 MHz, 144 MHz, and 164 MHz. All images are on the same colour scale. 
              }
         \label{three_freqs}}
   \end{figure*}
\end{appendix}




\end{document}